\documentclass[thmsa,11pt]{article}
\usepackage{amsfonts}
\usepackage{amsmath}
\usepackage{amssymb}
\usepackage{amsfonts}
\usepackage{amsmath}
\usepackage{epsfig,multicol}
\usepackage{graphicx}

\input{tcilatex}

\begin{document}

\setcounter{page}{0} \topmargin0pt \oddsidemargin5mm \renewcommand{%
\thefootnote}{\fnsymbol{footnote}} \newpage \setcounter{page}{0} 
\begin{titlepage}
\begin{flushright}
% EMPG\\
\end{flushright}
\vspace{0.5cm}
\begin{center}
{\Large {\bf PT Symmetry on the Lattice:\\ The Quantum Group Invariant XXZ Spin-Chain} }

\vspace{0.8cm}
{ \large Christian Korff$^1$ and Robert Weston$^2$}

\vspace{0.5cm}
{\em 
1) Department of Mathematics, University of Glasgow, \\
University Gardens, Glasgow G12 8QW, UK\\
\medskip
2) Department of Mathematics and the Maxwell Institute for Mathematical Sciences,\\
 Heriot-Watt University, Edinburgh EH14 4AS, UK}
\end{center}
\vspace{0.2cm}
 
\renewcommand{\thefootnote}{\arabic{footnote}}
\setcounter{footnote}{0}

\begin{abstract}
We investigate the PT-symmetry of the quantum group invariant XXZ chain. We 
show that the PT-operator commutes with the quantum group action and also discuss the 
transformation properties of the Bethe wavefunction. 
We exploit the fact that the Hamiltonian is an element of the Temperley-Lieb algebra in order
to give an explicit and exact construction of an operator that ensures quasi-Hermiticity 
of the model. 
This construction  relies on earlier ideas related to quantum group 
reduction. We then employ this result in connection with the quantum analogue of 
Schur-Weyl duality to introduce a dual pair of C-operators, both of which 
have closed algebraic expressions. These are novel, exact results connecting the 
research areas of integrable lattice systems and non-Hermitian Hamiltonians.

\medskip
\par\noindent
\end{abstract}
\vfill{ \hspace*{-9mm}
\begin{tabular}{l}
\rule{6 cm}{0.05 mm}\\
c.korff@maths.gla.ac.uk and r.a.weston@ma.hw.ac.uk
\end{tabular}}
\end{titlepage}\newpage 

\section{Introduction}

Recent years have seen growing interest in non-Hermitian Hamiltonians and
their interpretation in quantum mechanics (see \cite%
{Proc1,Proc2,Proc3,Proc4,Proc5} and references therein). There are several
motivations to study such systems. Non-Hermitian Hamiltonians appear to be
physically interesting: they can arise in connection with perturbative or
effective descriptions of physical phenomena. More generally, it has been
suggested that Hermiticity should not be the decisive criterion in deciding
whether the associated quantum mechanics system is physically sound, but
rather the reality of its spectrum; see the recent reviews \cite{Bender07} 
\cite{CFMFAF} and references therein. However, in order to ensure the
unitarity of the time evolution operator one is necessarily lead back to the
requirement of a Hermitian Hamiltonian. The way out is the introduction of a
new inner product on the Hilbert space of quantum states with respect to
which the Hamilton operator becomes Hermitian.

To be concrete, let us consider a Hamiltonian $H$ defined on a Hilbert space 
$\mathfrak{H}$ with inner product $\left\langle \cdot ,\cdot \right\rangle :%
\mathfrak{H\times H}\rightarrow \mathbb{C}$. Suppose $H$ is non-Hermitian
with respect to the inner product $\left\langle \cdot ,\cdot \right\rangle $%
, i.e. 
\begin{equation}
H\neq H^{\ast }
\end{equation}%
with $\ast $ denoting the Hermitian adjoint (or complex conjugate transpose
in the case of matrices). Provided there exists a \emph{self-adjoint,
invertible and positive}\noindent\ operator $\eta :\mathfrak{H}\rightarrow 
\mathfrak{H}$ such that%
\begin{equation}
\eta H=H^{\ast }\eta ,  \label{nHHn}
\end{equation}%
one can introduce another, different inner product 
\begin{equation}
\left\langle \cdot ,\cdot \right\rangle _{\eta }:\mathfrak{H}\times 
\mathfrak{H}\rightarrow \mathbb{C},\qquad \left\langle x,y\right\rangle
_{\eta }:=\left\langle x,\eta y\right\rangle  \label{etaprod}
\end{equation}%
with respect to which the Hamiltonian is now Hermitian,%
\begin{equation}
\left\langle x,Hy\right\rangle _{\eta }=\left\langle Hx,y\right\rangle
_{\eta },\qquad x,y\in \mathfrak{H}\;.
\end{equation}%
Since $\eta >0$, it possesses a unique positive square root which we denote
by $\eta ^{\frac{1}{2}}:\mathfrak{H}\rightarrow \mathfrak{H}$. Thus, one
might alternatively consider the Hamiltonian 
\begin{equation}
\tilde{H}=\eta ^{\frac{1}{2}}H\eta ^{-\frac{1}{2}}  \label{h}
\end{equation}%
which is Hermitian with respect to the original inner product $\left\langle
\cdot ,\cdot \right\rangle $, 
\begin{equation}
\tilde{H}^{\ast }=\eta ^{-\frac{1}{2}}H^{\ast }\eta ^{\frac{1}{2}}=\tilde{H}%
\ .
\end{equation}

Non-Hermitian Hamiltonians which allow for the existence of such a positive
map $\eta $ have been named \emph{quasi-Hermitian} in the literature, see 
\cite{SGH92}, \cite{Most04} and references therein. Note, that while the
property of quasi-Hermiticity obviously ensures the reality of the spectrum
of $H$, the converse is not true. In fact, in this article we will consider
a non-Hermitian Hamilton operator which for special values of a coupling
parameter has real spectrum but does not allow for a positive map $\eta $\
satisfying (\ref{nHHn}) unless a reduction of the state space is carried out
first.\smallskip

It is thus desirable for practical purposes to find a simple criterion which
allows one to decide whether a given non-Hermitian Hamiltonian might in fact
be quasi-Hermitian. Based on a large number of examples, it has been
suggested that such a criterion is $PT$-symmetry, namely the invariance of
the Hamiltonian under a simultaneous change of parity $P$ and time reversal $%
T$ (see the review \cite{Bender07} and references therein). While this has
proved to be an effective way of singling out many quasi-Hermitian Hamilton
operators within the pool of non-Hermitian ones, $PT$-symmetry of the
Hamiltonian alone is not mathematically sufficient to establish even the
reality of the spectrum due to the fact that time reversal is an anti-linear
operator \cite{Weig03}.

Nevertheless, in this paper we also investigate a Hamiltonian which is $PT$%
-symmetric. This symmetry is in our case even distinguished in light of an
underlying algebraic structure which we are going to exploit to establish
its quasi-Hermiticity. Thus, our example will confirm once more the
usefulness of $PT$-symmetry as a pre-selection tool for quasi-Hermitian
Hamilton operators.

Alternative approaches to constructing $\eta $ have been pursued in the
literature (for references we refer the reader to the reviews \cite{Bender07}%
, \cite{CFMFAF}). One is based on the explicit solution of the eigenvalue
problem of the Hamiltonian and through the construction of bi-orthonormal
systems. Other approaches rest on perturbation theory and the
Baker-Campbell-Hausdorff formula. Both methods have practical limitations,
in particular the former suffers from the fact that one seldom has exact
expressions for all the eigenfunctions. It is in this context that
integrable or exactly solvable systems play a special role as they allow one
in principle to obtain exact, non-perturbative information. Previous
applications of integrable systems in the context of non-Hermitian
Hamiltonians include for example the correspondence between integrable
systems and ordinary differential equations; see \cite{Dorey:2007zx} for a
recent review. One might also directly consider non-Hermitian deformations
of integrable systems respecting $PT$-symmetry; recent examples are
discussed in \cite{CMBDCBJCEF07,AF07}.\smallskip

In this paper, we shall consider a well-known integrable quantum Hamiltonian 
\cite{Alc87,PS90,JK94} which for certain values of a parameter $q$ is
non-Hermitian. The novel aspect here is that we introduce the concept of $PT$%
-symmetry and the so called $C\,$-operator for a discrete lattice model.
Namely, we are going to consider the quantum group invariant $XXZ$
spin-chain Hamiltonian%
\begin{equation}
H=\frac{1}{2}\sum_{i=1}^{N-1}\left\{ \sigma _{i}^{x}\sigma _{i+1}^{x}+\sigma
_{i}^{y}\sigma _{i+1}^{y}+\Delta _{+}~\left( \sigma _{i}^{z}\sigma
_{i+1}^{z}-1\right) \right\} +\Delta _{-}~\frac{(\sigma _{1}^{z}-\sigma
_{N}^{z})}{2}\ .  \label{H}
\end{equation}%
Here, the anisotropy parameters $\Delta _{\pm }$ are defined in terms of the
single variable $q$,%
\begin{equation}
\Delta _{\pm }=\frac{q\pm q^{-1}}{2}\ ,  \label{Delta}
\end{equation}%
and $\{\sigma _{i}^{x,y,z}\}$ denote the Pauli matrices acting on the $i^{%
\text{th}}$ site of the spin-chain represented by the state space $\mathfrak{%
H}=(\mathbb{C}^{2})^{\otimes N}$. We will focus on the case when the complex
parameter $q$ lies on the unit circle $\mathbb{S}^{1}$, this implies that $H$
is non-Hermitian, 
\begin{equation}
q\in \mathbb{S}^{1}:\qquad H\neq H^{\ast }\ .
\end{equation}%
This case of $q$ on the unit circle is of particular interest, since then
the corresponding lattice model is believed to correspond in the
thermodynamic limit to a conformal field theory with central charge \cite%
{Alc87,PS90}%
\begin{equation}
c=1-\frac{6}{(r-1)r},\qquad q=\exp \left( \frac{i\pi }{r}\right) \ .
\label{c}
\end{equation}%
When $r\in \mathbb{N},$ $r>2$, i.e. $q$ is a root of unity, the above
central charge value matches the one from the minimal unitary series.
Besides this connection with conformal field theory, which has fuelled
ongoing interest in this particular spin-chain Hamiltonian, there are
several algebras which play an important role in the eigenvalue problem of
this Hamiltonian. One is its quantum group invariance, the other is its
connection with the Hecke and Temperley-Lieb algebras. We will use the
representation theory of these algebras to obtain explicit and exact
expressions for $\eta $. To our knowledge there are only a few, simple
systems where this has been previously achieved; see the reviews \cite%
{Bender07}, \cite{CFMFAF}.\smallskip 

Our aim is to connect the discussion of quasi-Hermitian Hamilton operators
and $PT$-symmetry with a procedure known as \emph{quantum group reduction at
roots of unity} in the integrable lattice models community. While this
latter procedure was introduced a long time ago \cite{RS90}, its connection
with the more recent discussion of non-Hermitian Hamilton operators has not
been previously investigated. Moreover, we will derive for this particular
case novel expressions for $\eta $ in terms of the Hecke algebra, thus
giving a purely algebraic definition of the new Hilbert space structure in
which (\ref{H}) is Hermitian. We will also consider a special segment of the
unit circle where $q$ is not a root unity. For this case, no reduction of
the state space is required.\smallskip

The content of this paper is as follows. In Section 2, we recall the
relevant algebraic structures. We hope that this section will keep the paper
self-contained and make it more accessible to a wider audience, in
particular the community working on non-Hermitian Hamilton operators. In
Section 3, we introduce the concept of $PT$-symmetry on the lattice,
explaining its connection with quasi-Hermiticity. This exposition is aimed
primarily at researchers in the field of integrable systems. We go on to
discuss the relation of the $P$ and $T$ operators with the action of our two
algebras. We also give the action of $PT$ on the Bethe wave function. In
Section 4, we give a construction of $\eta $ in terms of the path basis for $%
q$ a root of unity (see equation (\ref{eta})). In Section 5, we define a $C$
operator as $C=P\eta $ and give a simple, purely algebraic realization of it
in terms of a certain braid operator associated with the Hecke algebra. We
also define another operator $C^{\prime }$ that has properties that are
similar to $C$ and yet is loosely speaking "dual" to it. In particular, $C$
is an element of the Hecke algebra that commutes with the quantum group,
while $C^{\prime }$ is element of the quantum group that commutes with the
Hecke algebra. Finally, we make some concluding comments in Section 6.

\section{The XXZ chain and its related algebras}

As mentioned in the introduction, we will exploit several algebraic
structures associated with the spin-chain Hamiltonian in order to establish
its quasi-Hermiticity. In this section, we review the definition of these
algebras and their relation to the Hamiltonian to keep this article
self-contained. For details concerning the definitions of the various
algebras and their properties we refer the reader to e.g. \cite{chpr94}.

\subsection{The Temperley-Lieb and Hecke algebra}

To make contact with the first algebra, we rewrite the spin-chain
Hamiltonian (\ref{H}) in terms of "local" operators which only contain
nearest-neighbour interactions,%
\begin{equation}
H=\sum_{i=1}^{N-1}E_{i},\qquad E_{i}=\frac{\sigma _{i}^{x}\sigma
_{i+1}^{x}+\sigma _{i}^{y}\sigma _{i+1}^{y}}{2}+\Delta _{+}\frac{\sigma
_{i}^{z}\sigma _{i+1}^{z}-1}{2}+\Delta _{-}\frac{\sigma _{i}^{z}-\sigma
_{i+1}^{z}}{4}\ .  \label{TLrep}
\end{equation}%
Let $V=\mathbb{C}^{2}$ be a two-dimensional complex vector space, then these
"local Hamiltonians" $E_{i}$ provide a particular representation%
\begin{equation}
\pi _{TL}:TL_{N}(q)\rightarrow \limfunc{End}V^{\otimes N},\qquad
e_{i}\mapsto \pi _{TL}(e_{i}):=E_{i}  \label{piTL}
\end{equation}%
of an abstract algebra $TL_{N}(q)$, known as the Temperley-Lieb
algebra.\smallskip

\noindent \underline{\textsc{Definition 2.1}}\textsc{\ [Temperley-Lieb
Algebra].} \emph{The Temperley-Lieb algebra }$TL_{N}(q)$\emph{\ is obtained
from }$N-1$\emph{\ generators }$\{e_{1},e_{2},...,e_{N-1}\}$\emph{\
satisfying the commutation relations}%
\begin{eqnarray}
e_{i}^{2} &=&-(q+q^{-1})e_{i},  \notag \\
e_{i}e_{i\pm 1}e_{i} &=&e_{i},  \notag \\
e_{i}e_{j} &=&e_{j}e_{i},\qquad |i-j|>1\;.  \label{TLdef}
\end{eqnarray}

\noindent In the present context, the Temperley-Lieb algebra plays the role
of a spectrum generating algebra. Note that the representation defined by (%
\ref{TLrep}) is non-Hermitian for $q$ on the unit circle. This is most
easily seen when expressing the Temperley-Lieb generators as 4 by 4 matrices
acting on the $i^{\,\text{th}}$ and ($i+1$)$^{\text{th}}$ factor in the
spin-chain, 
\begin{equation}
E_{i}=\left( 
\begin{array}{cccc}
0 & 0 & 0 & 0 \\ 
0 & -q^{-1} & 1 & 0 \\ 
0 & 1 & -q & 0 \\ 
0 & 0 & 0 & 0%
\end{array}%
\right) _{i,i+1}\neq \bar{E}_{i}=E_{i}^{\ast }\;.  \label{TLspin}
\end{equation}%
The irreducible representations of the Temperley-Lieb algebra will be
discussed in subsequent sections of this paper and in Appendix A.

The Temperley-Lieb algebra is closely related to another algebra which can
be thought of as a generalisation or $q$-deformation of the group algebra of
the symmetric group.$\mathfrak{\smallskip }$

\noindent \underline{\textsc{Definition 2.2}}\textsc{\ [Hecke Algebra].} 
\emph{The Hecke algebra }$H_{N}(q)$\emph{\ is obtained from }$N-1$\emph{\
generators }$\{b_{i}\}_{i=1}^{N-1}$\emph{\ obeying the defining relations,}%
\begin{eqnarray}
b_{i}b_{i}^{-1} &=&b_{i}^{-1}b_{i}=1,  \notag \\
b_{i}b_{i+1}b_{i} &=&b_{i+1}b_{i}b_{i+1},  \notag \\
b_{i}b_{j} &=&b_{j}b_{i},\qquad |i-j|>1  \label{Braid def}
\end{eqnarray}%
\emph{and the quadratic relation}%
\begin{equation}
(b_{i}+q)(b_{i}-q^{-1})=0\ .  \label{Hecke square}
\end{equation}

\noindent $H_{N}(q)$ is the group algebra of Artin's braid group factored by
the relation (\ref{Hecke square}). Hecke and Temperley-Lieb algebras are
related by the following surjective homomorphism,%
\begin{equation}
\varphi :H_{N}(q)\rightarrow TL_{N}(q):\text{\quad }b_{i}\mapsto q^{-1}+e_{i}%
\text{\quad\ and\quad\ }b_{i}^{-1}\mapsto q+e_{i}\ .  \label{hom}
\end{equation}%
Using this homomorphism, we extend the representation of the Temperley-Lieb
algebra (\ref{TLrep}) to the Hecke algebra%
\begin{equation}
\pi _{H}:H_{N}(q)\rightarrow \limfunc{End}V^{\otimes N},\qquad b_{i}\mapsto
\pi _{H}(b_{i}):=(\pi _{TL}\circ \varphi )(b_{i})  \label{piH}
\end{equation}%
and obtain%
\begin{equation}
b_{i}\mapsto \pi _{H}(b_{i}):=B_{i}=\left( 
\begin{array}{cccc}
q^{-1} & 0 & 0 & 0 \\ 
0 & 0 & 1 & 0 \\ 
0 & 1 & q^{-1}-q & 0 \\ 
0 & 0 & 0 & q^{-1}%
\end{array}%
\right) _{i,i+1}\ .  \label{Hecke spin}
\end{equation}%
Our motivation to introduce the Hecke algebra will become clear when
constructing the new Hilbert space structure with respect to which the
Hamiltonian (\ref{H}) becomes Hermitian. In a purely algebraic context, the
Hecke algebra is closely connected to the intertwiners of the quantum group $%
U_{q}(sl_{2})$ which describes the degenerate eigenspaces of the Hamiltonian
(\ref{H}).

\subsection{Quantum group invariance}

The spin-chain Hamiltonian (\ref{H}) possesses several symmetries. In the
present context, the most distinguished is its quantum group invariance,
which we shall employ in order to define a new Hilbert space structure in
which $H$ becomes Hermitian.\smallskip

\noindent \underline{\textsc{Definition 2.3}}\textsc{\ [Quantum Group].} 
\emph{The quantum group }$U_{q}(sl_{2})$\emph{\ is a quasi-triangular Hopf
algebra generated by the Chevalley-Serre generators }$\{s^{\pm },q^{\pm
s^{z}}\}$\emph{. The latter are subject to the commutation relations}%
\begin{equation}
q^{s^{z}}q^{-s^{z}}=q^{-s^{z}}q^{s^{z}}=1,\qquad q^{s^{z}}s^{\pm
}q^{-s^{z}}=q^{\pm 1}s^{\pm },\qquad \lbrack s^{+},s^{-}]=[2s^{z}]_{q}:=%
\frac{q^{2s^{z}}-q^{-2s^{z}}}{q-q^{-1}}\ .
\end{equation}%
\emph{The Hopf algebra structure includes the notions of coproduct }$\Delta
:U_{q}(sl_{2})\rightarrow U_{q}(sl_{2})\otimes U_{q}(sl_{2})$\emph{, }%
\begin{equation}
\Delta (s^{\pm })=q^{s^{z}}\otimes s^{\pm }+s^{\pm }\otimes
q^{-s^{z}},\qquad \Delta (q^{\pm s^{z}})=q^{\pm s^{z}}\otimes q^{\pm s^{z}},
\label{cop}
\end{equation}%
\emph{antipode }$\gamma :U_{q}(sl_{2})\rightarrow U_{q}(sl_{2})$\emph{\ }%
\begin{equation}
\gamma (q^{\pm s^{z}})=q^{\mp s^{z}},\qquad \gamma (s^{\pm })=-q^{\pm
1}s^{\pm },  \label{antip}
\end{equation}%
\emph{and co-unit }$\varepsilon :U_{q}(sl_{2})\rightarrow \mathbb{C},$%
\begin{equation}
\varepsilon (q^{\pm s^{z}})=1,\qquad \varepsilon (s^{\pm })=0\;.  \label{co1}
\end{equation}

\noindent In the following, we will work with the familiar two-dimensional
spin-1/2 representation $\pi :U_{q}(sl_{2})\rightarrow \limfunc{End}V,\;V=%
\mathbb{C}^{2},$ in terms of Pauli matrices, i.e. 
\begin{equation}
\pi :\quad s^{\pm }\mapsto \sigma ^{\pm }\qquad \text{and\qquad }q^{\pm
s^{z}}\mapsto q^{\pm \sigma ^{z}}\;.
\end{equation}%
The action of the quantum group generators on the spin-chain $V^{\otimes N}$
is then obtained via successive application of the coproduct by defining $%
\Delta _{n}=(\Delta \otimes 1)\Delta _{n-1}$ starting with $\Delta
_{2}\equiv \Delta $,%
\begin{equation}
\pi _{U}:U_{q}(sl_{2})\rightarrow \limfunc{End}V^{\otimes N},\qquad x\mapsto
\pi ^{\otimes N}(\Delta _{N}(x)),\quad N>1\;.  \label{piU}
\end{equation}%
We shall denote the images of the Chevalley-Serre generators $\{s^{\pm
},q^{\pm s^{z}}\}$ under $\pi _{U}$ by capital letters. They read explicitly%
\begin{equation}
q^{\pm s^{z}}\mapsto \pi _{U}(q^{\pm s^{z}})\equiv q^{\pm S^{z}},\qquad
S^{z}=\frac{1}{2}\tsum_{i=1}^{N}\sigma _{i}^{z}  \label{Sz}
\end{equation}%
and%
\begin{equation}
s^{\pm }\mapsto \pi _{U}(s^{\pm })\equiv S^{\pm }=\sum_{i=1}^{N}q^{\frac{%
\sigma ^{z}}{2}}\otimes \cdots \otimes \underset{i^{\text{th}}}{\sigma ^{\pm
}}\otimes q^{-\frac{\sigma ^{z}}{2}}\cdots \otimes q^{-\frac{\sigma ^{z}}{2}%
}\ .  \label{Spm}
\end{equation}%
For $q$ on the unit circle, these generators are non-Hermitian. In fact, one
has the relation%
\begin{equation}
\left( S^{\pm }\right) ^{\ast }=S_{\text{op}}^{\mp }\equiv \pi ^{\otimes
N}(\Delta _{N}^{\text{op}}(s^{\mp }))  \label{Spmop}
\end{equation}%
where $S_{\text{op}}^{\mp }$ are the quantum group generators associated
with the opposite coproduct,%
\begin{equation}
\Delta ^{\text{op}}(s^{\pm })=q^{-s^{z}}\otimes s^{\pm }+s^{\pm }\otimes
q^{s^{z}}\ .  \label{opcop}
\end{equation}%
We shall refer to the associated Hopf algebra as $U_{q}^{\text{op}}(sl_{2})$%
. The permuted Hecke algebra generators $r_{i}=\tau _{i}b_{i},$ with $\tau
_{i}$ the permutation operator in the $i^{\text{th}}$ and ($i+1$)$^{\text{th}%
}$ factors, relate the two coproduct structures (\ref{cop}) and (\ref{opcop}%
),%
\begin{equation}
r_{i}(\sigma _{i}^{\pm }\otimes q^{-\sigma _{i+1}^{z}}+q^{\sigma
_{i}^{z}}\otimes \sigma _{i+1}^{\pm })=(\sigma _{i}^{\pm }\otimes q^{\sigma
_{i+1}^{z}}+q^{-\sigma _{i}^{z}}\otimes \sigma _{i+1}^{\pm })r_{i}\;.
\end{equation}%
The matrix $r_{i}$ is referred to as a quantum group intertwiner due to the
above relation; it is also commonly called the \textquotedblleft
R-matrix\textquotedblright . Both versions of the quantum group appear in
the present context: $U_{q}(sl_{2})$ is the symmetry algebra of the
Hamiltonian (\ref{H}), and $U_{q}^{\text{op}}(sl_{2})$ is the symmetry
algebra of its Hermitian adjoint $H^{\ast }$,%
\begin{equation}
\lbrack H,\pi _{U}(U_{q}(sl_{2}))]=[H^{\ast },\pi _{U}(U_{q}^{\text{op}%
}(sl_{2}))]=0\ .  \label{QGinv}
\end{equation}%
These commutation relations are a direct consequence of the following
relation which is a quantum analogue of the Schur-Weyl duality.$\mathfrak{%
\smallskip }$

\noindent \underline{\textsc{Theorem 2.1}} [Jimbo]\cite{Jim86b}\textsc{.} 
\emph{Let }$\pi _{U}:U_{q}(sl_{2})\rightarrow \limfunc{End}V^{\otimes N}$ 
\emph{and }$\pi _{H}:H_{N}(q)\rightarrow \limfunc{End}V^{\otimes N}$\emph{\
be the representations defined in (\ref{piU}) and (\ref{piTL}). Denote by }$%
\mathcal{U}^{\prime }$\emph{\ and }$\mathcal{H}^{\prime }$\emph{\ the
commutants of the operator algebras }$\mathcal{U}=\pi _{U}(U_{q}(sl_{2}))$%
\emph{\ and }$\mathcal{H}=\pi _{H}(H_{N}(q))$ \emph{in }$\limfunc{End}%
V^{\otimes N}$\emph{, respectively. Then we can identify}%
\begin{equation}
\mathcal{U}^{\prime }=\mathcal{H}\quad \text{\emph{and}}\quad \mathcal{H}%
^{\prime }\ =\mathcal{U}.  \label{QBFS duality}
\end{equation}%
$\mathfrak{\smallskip }$

\noindent In the limit $q\rightarrow 1$ this gives the familiar Schur-Weyl
duality with respect to the symmetric group. Note that in the simple case of 
$sl_{2}$ and $V=\mathbb{C}^{2}$ considered here, we can specialize in (\ref%
{QBFS duality}) to the Temperley-Lieb algebra. This is due to the fact that
for the local spin 1/2 representation (\ref{Hecke spin}), and only for this
representation, the homomorphism (\ref{hom}) in Theorem 2.1 can be
"inverted", i.e. the relation $E_{i}=q^{-1}-B_{i}$ yields a representation
of the Temperley-Lieb algebra.

The first relation in (\ref{QGinv}) now follows immediately from Theorem
2.1, and the second is obtained when taking the Hermitian adjoint and
employing $H(q)^{\ast }=H(q^{-1})$ together with (\ref{Spmop}). Below we
will make further use of the duality (\ref{QBFS duality}) when discussing
the quasi-Hermiticity properties of the Hamiltonian (\ref{H}) and when
introducing a new Hilbert space structure with respect to which $H$ is
Hermitian. This concludes our brief review of the algebraic structures
relevant for our discussion.

\section{PT symmetry on the lattice}

We now discuss $PT$-invariance for the spin-chain by introducing parity and
time reversal operators on the lattice. Their action is then related to the
quantum group invariance of the Hamiltonian and to the transformation
properties of a discrete wave function $\psi $.

\subsection{Definitions}

The Hamiltonian (\ref{H}) acts on the $N$-fold tensor product of a
two-dimensional space $V=\mathbb{C}^{2}$ with orthonormal basis vectors $v_{%
\frac{1}{2}}=\dbinom{1}{0}$ and $v_{-\frac{1}{2}}=\dbinom{0}{1}$. We then
have the following orthonormal basis in the Hilbert space $\mathfrak{H}%
=V^{\otimes N}$,%
\begin{equation}
\{\left\vert \alpha _{1},...,\alpha _{N}\right\rangle \equiv v_{\alpha
_{1}}\otimes \cdots \otimes v_{\alpha _{N}},~\alpha _{i}=\pm 1/2\}\ .
\label{spin-basis}
\end{equation}%
$\mathfrak{\smallskip }$

\noindent \underline{\textsc{Definition 3.1}}\textsc{\ [parity and
time-reversal operator].} \emph{On the above set of basis vectors we define
the linear operator }$P$\emph{\ by setting}%
\begin{equation}
P\left\vert \alpha _{1},...,\alpha _{N}\right\rangle =\left\vert \alpha
_{N},\alpha _{N-1},...,\alpha _{1}\right\rangle \ .  \label{P}
\end{equation}%
\emph{In contrast, the operator }$T$\emph{\ acts on the basis vectors as the
identity,} 
\begin{equation}
T\left\vert \alpha _{1},...,\alpha _{N}\right\rangle =\left\vert \alpha
_{1},...,\alpha _{N}\right\rangle ,  \label{T}
\end{equation}%
\emph{but is defined to be antilinear, such that} 
\begin{equation}
T\,\lambda \left\vert \alpha _{1},...,\alpha _{N}\right\rangle =\bar{\lambda}%
\left\vert \alpha _{1},...,\alpha _{N}\right\rangle ,\quad \lambda \in 
\mathbb{C}.
\end{equation}%
\emph{Thus, any matrix }$A$\emph{\ (such as the Hamiltonian }$A=H$\emph{) is
transformed into its complex conjugate under the adjoint action of }$T$,%
\begin{equation}
TAT=\bar{A}\ .
\end{equation}%
$\mathfrak{\smallskip }$

\noindent Note, that in the particular case considered here the Hamiltonian
is symmetric, $H^{t}=H$, and we therefore have the identity $\bar{H}=H^{\ast
}$. Together with the crucial relation 
\begin{equation}
PHP=H^{\ast },  \label{PHP}
\end{equation}%
which follows simply from the definitions (\ref{H}) and (\ref{P}), we obtain
as a consequence%
\begin{equation}
\lbrack PT,H]=0\ .  \label{PT}
\end{equation}%
Thus, the quantum group invariant XXZ Hamiltonian (\ref{H}) is $PT$%
-invariant. However, only if all the eigenfunctions can be chosen to be
simultaneous eigenfunctions of the $PT$-operator can one conclude that the
spectrum of $H$ must be real.

\subsection{The adjoint $PT$ action on the quantum group and Temperley-Lieb
algebra}

%Notice that the $PT$-symmetry is compatible with the quantum group symmetry
%as the Chevalley-Serre generators are also $PT$-invariant. 

The Chevalley-Serre generators have the following behaviour under the
adjoint action of $P$ and $T$ 
\begin{equation}
PS^{\pm }P=TS^{\pm }T=\sum_{i=1}^{N}q^{-\frac{\sigma ^{z}}{2}}\otimes \cdots
q^{-\frac{\sigma ^{z}}{2}}\otimes \underset{i^{\text{th}}}{\sigma ^{\pm }}%
\otimes q^{\frac{\sigma ^{z}}{2}}\otimes \cdots q^{\frac{\sigma ^{z}}{2}}=S_{%
\text{op}}^{\pm }\ .  \label{PSP}
\end{equation}%
Thus, the action of $PT$ commutes with that of the quantum group, and as
such $PT$ is distinguished from other symmetries of the Hamiltonian. For
example, we could have employed the spin-reversal operator%
\begin{equation}
R=\prod_{n=1}^{N}\sigma _{n}^{x}\   \label{R}
\end{equation}%
which leads to additional $PR$ and $TR$-symmetries,%
\begin{equation}
RHR=H^{\ast }\qquad \text{and}\qquad \lbrack PR,H]=[TR,H]=0\ .  \label{RHR}
\end{equation}%
Notice, however, that the latter do not commute with the quantum group
generators,%
\begin{equation}
RS^{\pm }R=S_{\text{op}}^{\mp },\quad \quad PR~S^{\pm }~PR=RT~S^{\pm
}~RT=S^{\mp }\ .  \label{RSR}
\end{equation}

The Temperley-Lieb algebra generators on the other hand are not $PT$%
-invariant. In the representation (\ref{TLrep}) they transform according to%
\begin{equation}
PE_{k}P=E_{N-k}^{\ast }=TE_{N-k}T\qquad \text{and so\qquad }%
PT~E_{k}=E_{N-k}~PT\;.  \label{PTLP}
\end{equation}%
For the spin-reversal operator we obtain the identities%
\begin{equation}
R~E_{k}R=E_{k}^{\ast },\quad PR~E_{k}PR=E_{N-k},\quad RT~E_{k}RT=E_{k}\ .
\label{RTLR}
\end{equation}%
Note that the $PR$ symmetry is insufficient to introduce a new Hilbert space
structure: $PR$ is not a positive operator, whence we cannot use it to
define an alternative inner product. In order to arrive at the latter we
need the $\eta $-operator which we discuss below. Let us summarize the
various transformation properties of the respective algebra generators in
the following table:\medskip

\begin{center}
\begin{tabular}{||c||c|c|}
\hline\hline
Operator & Temperley-Lieb & Quantum Group \\ \hline\hline
Parity reversal & $PE_{k}P=E_{N-k}^{\ast }$ & $PS^{\pm }P=S_{\text{op}}^{\pm
}$ \\ \hline
Time reversal & $TE_{k}T=E_{k}^{\ast }$ & $TS^{\pm }T=S_{\text{op}}^{\pm }$
\\ \hline
Spin reversal & $R~E_{k}R=E_{k}^{\ast }$ & $RS^{\pm }R=S_{\text{op}}^{\mp }$
\\ \hline
\end{tabular}%
\medskip

Table 1. Commutation relations.\medskip
\end{center}

\noindent In our subsequent discussion we will make frequent use of these
commutation relations.

\subsection{$PT$ symmetry and Bethe's wave function}

The term $PT$ symmetry is used in two senses in the quantum mechanics
literature: in the weak sense it means simply that $[PT,H]=0$; in the strong
sense the term means that in addition $PT|\psi \rangle \propto |\psi \rangle 
$ for all eigenvectors $|\psi \rangle $ of $H$. Clearly, this latter
property does not follow from $[PT,H]=0$. In particular if the energy
eigenvalue of $|\psi \rangle $ is complex, then the antilinearity of $T$
means that the eigenvalue of $PT|\psi \rangle $ is its complex conjugate. If
a system displays $PT$ symmetry in the weak but not the strong sense, then
it is said to display spontaneous breaking of $PT$ symmetry \cite{Weig03}.

In this subsection, we discuss the transformation properties of the
eigenvectors of the Hamiltonian $H$ under $PT$-symmetry. To this end, the
formalism of the coordinate Bethe ansatz turns out to be most convenient.
This will allow us to determine whether there is a spontaneous breakdown of $%
PT$-symmetry and further motivate our existing definition of the parity and
time reversal operators by relating them to the transformation behaviour of
a "discrete wavefunction" describing the Bethe vectors.\smallskip

The coordinate Bethe ansatz for the Hamiltonian (\ref{H}) has been
previously discussed in the literature \cite{Alc87,PS90} and we refer the
reader to these works for the details of the derivation of the Bethe ansatz
equations. We emphasize, however, that $PT$-symmetry has not been discussed
in these works and this is the novel aspect which we want to highlight
here.\smallskip

Before we review the coordinate Bethe ansatz in the context of $PT$%
-symmetry, we first introduce the $PT$-action on discrete wave functions.
Quite generally, any vector $\left\vert \psi \right\rangle $ in the Hilbert
space $\mathfrak{H}=V^{\otimes N}$ with $n$ down-spins is in one-to-one
correspondence with a discrete wave function $\psi (x_{1},...,x_{n})$
according to the relation%
\begin{equation}
\left\vert \psi \right\rangle =\sum_{1\leq x_{1}<...<x_{n}\leq N}\psi
(x_{1},...,x_{n})\sigma _{x_{1}}^{-}\cdots \sigma _{x_{n}}^{-}\left\vert 
\tfrac{1}{2},...,\tfrac{1}{2}\right\rangle ,  \label{psi}
\end{equation}%
where $|\psi (x_{1},...,x_{n})|^{2}$ can be interpreted as the probability
to find the $n$ down-spins located at the lattice sites $x_{1},...,x_{n}$.
According to our previous definitions (\ref{P}), (\ref{T}) the
transformation behaviour of the wave function under parity-reversal is%
\begin{equation}
\psi (x_{1},...,x_{n})\overset{P}{\rightarrow }\psi
(N+1-x_{n},...,N+1-x_{1})\ .  \label{Ppsi}
\end{equation}%
Time reversal simply amounts to complex conjugation,%
\begin{equation}
\psi (x_{1},...,x_{n})\overset{T}{\rightarrow }\bar{\psi}(x_{1},...,x_{n})\ .
\label{Tpsi}
\end{equation}%
These two transformation properties are clearly analogues of the
transformation properties of a standard \emph{continuous} wave function
describing a many-particle system confined to a finite interval on the real
line.

The eigenvectors of the Hamiltonian (\ref{H}) corresponding to highest
weight vectors with respect to the quantum group symmetry (\ref{QGinv}) can
be described in terms of Bethe's wave function, $\psi =\psi _{\boldsymbol{k}%
} $, which can be interpreted as a superposition of reflected plane waves
with quasi-momenta $\boldsymbol{k}=(k_{1},...,k_{n})$, and is defined by%
\begin{equation}
\psi _{\boldsymbol{k}}(x_{1},...,x_{n})=\sum_{p,\varepsilon
}(-1)^{|p|}A(\varepsilon _{1}k_{p_{1}},\cdots ,\varepsilon
_{n}k_{p_{n}})e^{i(\varepsilon _{1}k_{p_{1}}x_{1}+\cdots +\varepsilon
_{n}k_{p_{n}}x_{n})}~.  \label{Bethepsi}
\end{equation}%
The sum runs over all permutations $p=(p_{1},...,p_{n})\in S_{n}$ of the
index set $(1,...,n)$ as well as all possible sign changes $\varepsilon
=(\varepsilon _{1},...,\varepsilon _{n}),\;\varepsilon _{i}=\pm 1$ (due to
the reflection at the boundaries of the finite spin-chain). The symbol $|p|$
indicates the sign $\pm 1$ of the permutation, with the choice $|(1,2,\cdots
,n)|=+1$. It is shown in \cite{Alc87} that the coefficients $A(k_{1},\cdots
,k_{n})$ have the form%
\begin{equation}
A(k_{1},\cdots ,k_{n})=\prod_{j}\beta
(-k_{j})\prod_{j<l}B(-k_{j},k_{l})e^{-ik_{l}},  \label{amp}
\end{equation}%
where we have introduced the functions 
\begin{equation}
\beta (k)=(1-qe^{-ik})e^{i(N+1)k}
\end{equation}%
and%
\begin{equation}
B(k_{1},k_{2})=s(-k_{1},k_{2})s(k_{2},k_{1}),\qquad
s(k_{1},k_{2}):=(1-(q+q^{-1})e^{ik_{1}}+e^{i(k_{1}+k_{2})})\ .
\end{equation}%
For later use we note the simple identities (note that we have $\bar{q}%
=q^{-1}$) 
\begin{equation}
\overline{\beta (-\bar{k})}=-q^{-1}e^{i(2N+1)k}\beta (-k),\quad \overline{B(-%
\bar{k}_{1},\bar{k}_{2})}=B(-k_{2},k_{1})e^{-i(k_{1}+k_{2})}\frac{%
s(-k_{2},k_{1})}{s(k_{1},-k_{2})}  \label{idbethe}
\end{equation}%
In order that $|\psi _{\boldsymbol{k}}\rangle $ be an eigenvector, it is
necessary that the quasi-momenta $k_{1},\cdots ,k_{n}$ satisfy the Bethe
ansatz equations%
\begin{equation}
e^{2iNk_{j}}=\prod_{l\neq j}\frac{B(-k_{j},k_{l})}{B(k_{j},k_{l})}\ .
\label{BAE}
\end{equation}%
The corresponding eigenvalue of the Hamiltonian is%
\begin{equation}
\lambda =(N-1)\Delta _{+}+4\sum_{j=1}^{n}\left( \cos k_{j}-\Delta
_{+}\right) \ .  \label{Bethespec}
\end{equation}%
The analytic solutions to the Bethe ansatz equations are not known. However,
we can discuss in abstract terms the action of parity and time reversal on
the Bethe wave function. We have 
\begin{multline}
PT\psi _{\boldsymbol{k}}(x_{1},\cdots ,x_{n})=\sum_{p,\varepsilon }(-1)^{|p|}%
\overline{A(\varepsilon _{1}k_{p_{1}},\cdots ,\varepsilon _{n}k_{p_{n}})}%
e^{-i(\varepsilon _{1}\bar{k}_{p_{1}}(N+1-x_{n})+\cdots +\varepsilon _{n}%
\bar{k}_{p_{n}}(N+1-x_{1}))}  \notag \\
=-\sum_{p,\varepsilon }(-1)^{|p|}\overline{A(\varepsilon
_{n}k_{p_{n}},\cdots ,\varepsilon _{1}k_{p_{1}})}e^{-i(N+1)(\varepsilon _{1}%
\bar{k}_{p_{1}}+\cdots +\varepsilon _{n}\bar{k}_{p_{n}})}e^{i(\varepsilon
_{1}\bar{k}_{p_{1}}x_{1}+\cdots +\varepsilon _{n}\bar{k}_{p_{n}}x_{n})}
\end{multline}%
The minus sign comes from the sign difference of the two permutations $%
(p_{1},\cdots ,p_{n})$ and $(p_{n},\cdots ,p_{1})$. Now let us assume that
the Bethe roots $k_{i}$ consist of $m$ complex pairs $(k_{i},k_{i^{\prime
}}) $ of the form $\bar{k}_{i}=\pm k_{i^{\prime }}$, and $n-2m$ real roots.
This assumption is certainly consistent with the reality of the spectrum due
to (\ref{Bethespec}). Under this assumption, we can write 
\begin{multline}
PT\psi _{\boldsymbol{k}}(x_{1},\cdots ,x_{n})= \\
(-)^{1+m}\sum_{p,\varepsilon }(-1)^{|p|}\overline{A(\varepsilon _{n}\bar{k}%
_{p_{n}},\cdots ,\varepsilon _{1}\bar{k}_{p_{1}})}e^{-i(N+1)(\varepsilon
_{1}k_{p_{1}}+\cdots +\varepsilon _{n}k_{p_{n}})}e^{i(\varepsilon
_{1}k_{p_{1}}x_{1}+\cdots +\varepsilon _{n}k_{p_{n}}x_{n})}.
\end{multline}%
It follows that we have $PT\psi _{\boldsymbol{k}}(x_{1},\cdots
,x_{n})\propto \psi _{\boldsymbol{k}}(x_{1},\cdots ,x_{n})$ if 
\begin{equation}
A(k_{1},\cdots ,k_{n})\propto \overline{A(\bar{k}_{n},\cdots ,\bar{k}_{1})}%
e^{-i(N+1)(k_{1}+\cdots k_{n})}
\end{equation}%
for any set of $\boldsymbol{k}=(k_{1},\cdots ,k_{n})$ satisfying the Bethe
ansatz equations (\ref{BAE}). This is a consequence of the following
proposition.$\mathfrak{\smallskip }$

\noindent \underline{\textsc{Proposition 3.1}} \textsc{.} \emph{If }$%
k_{1},\cdots ,k_{n}$\emph{\ satisfy the Bethe equations (\ref{BAE}), we have}
\begin{equation}
\overline{A(\bar{k}_{1},\cdots ,\bar{k}_{n})}=(-q)^{-n}e^{i(N+1)(k_{1}+%
\cdots +k_{n})}A(k_{n},\cdots ,k_{1})\ .
\end{equation}%
$\mathfrak{\smallskip }$

\noindent \textsc{Proof.} Using the above identities (\ref{idbethe}) for $%
\bar{\beta}$ and $\bar{B}$, we have 
\begin{equation}
\overline{A(\bar{k}_{1},\cdots ,\bar{k}_{n})}=(-q)^{-n}e^{i(2N+1)(k_{1}+%
\cdots +k_{n})}\prod_{j}\beta (-k_{j})\prod_{j<l}B(-k_{l},k_{j})e^{-ik_{j}}%
\frac{s(-k_{l},k_{j})}{s(k_{j},-k_{l})}  \label{ex1}
\end{equation}%
The next step is to take the product of the Bethe equations over all $%
j=1,\cdots ,n$. This gives 
\begin{eqnarray}
e^{2iN(k_{1}+\cdots +k_{n})} &=&\prod_{j}\prod_{l\neq j}\frac{%
s(k_{j},k_{l})s(k_{l},-k_{j})}{s(-k_{j},k_{l})s(k_{l},k_{j})}%
=\prod_{j}\prod_{l\neq j}\frac{s(k_{l},-k_{j})}{s(-k_{j},k_{l})},  \notag \\
&=&\prod_{j<l}\frac{s(k_{l},-k_{j})}{s(-k_{j},k_{l})}\prod_{j<l}\frac{%
s(k_{j},-k_{l})}{s(-k_{l},k_{j})}=\prod_{j<l}\frac{s^{2}(k_{j},-k_{l})}{%
s^{2}(-k_{l},k_{j})},
\end{eqnarray}%
where we have used the identity $%
s(k_{1},k_{2})=e^{i(k_{1}+k_{2})}s(-k_{2},-k_{1})$ in the last step. Then we
have 
\begin{equation}
\prod_{j<l}\frac{s(-k_{l},k_{j})}{s(k_{j},-k_{l})}=e^{-iN(k_{1}+\cdots
+k_{n})}
\end{equation}%
The left-hand-side of this expression appears in Equation (\ref{ex1}), and
after substituting in we arrive at 
\begin{eqnarray}
\overline{A(\bar{k}_{1},\cdots ,\bar{k}_{n})} &=&(-q)^{-n}e^{i(N+1)(k_{1}+%
\cdots +k_{n})}\prod_{j}\beta (-k_{j})\prod_{j<l}B(-k_{l},k_{j})e^{-ik_{j}} 
\notag \\
&=&(-q)^{-n}e^{i(N+1)(k_{1}+\cdots +k_{n})}A(k_{n},\cdots ,k_{1})
\end{eqnarray}%
\hfill {$\blacksquare $}

If follows that for each Bethe vector with $n$ down spins (i.e. $S_{z}$
eigenvalue $N/2-n$), $m$ complex pairs $(k_{i},k_{i^{\prime }})$ of the form 
$\bar{k}_{i}=\pm k_{i^{\prime }}$ and $n-2m$ real roots, we have 
\begin{equation}
PT\psi _{\boldsymbol{k}}(x_{1},\cdots ,x_{n})=(-1)^{1+m}(-q)^{-n}\psi _{%
\boldsymbol{k}}(x_{1},\cdots ,x_{n})  \label{Ptbethepsi}
\end{equation}%
and hence 
\begin{equation}
PT|\psi _{\boldsymbol{k}}\rangle =(-1)^{1+m}(-q)^{-n}|\psi _{\boldsymbol{k}%
}\rangle \ .  \label{PTbethe}
\end{equation}%
Thus, for this particular set of eigenvectors at least, and - because of (%
\ref{PSP}) - for their associated degenerate eigenspaces, we can conclude
that $PT$-symmetry is not spontaneously broken. However, since it is
generally difficult to verify the precise nature of the Bethe roots (i.e.
whether only real or complex pairs of them occur) as well as to rigorously
prove that the Bethe ansatz yields the complete set of eigenvectors,
alternative arguments have to be employed to prove the reality of the
spectrum of (\ref{H}).

\section{Exact construction of the quasi-Hermiticity operator $\protect\eta $%
}

As outlined in the introduction, we need to find a positive definite,
Hermitian and invertible operator $\eta $ satisfying (\ref{nHHn}) in order
to define a new Hilbert space structure (\ref{etaprod}) with respect to
which the Hamiltonian (\ref{H}) becomes Hermitian. For many non-Hermitian
systems in the literature, this has been achieved by using bi-orthonormal
systems of Hamiltonian eigenvectors. While we will not follow this approach
in our construction of $\eta $, because the solutions of the Bethe ansatz
equations (\ref{BAE}) are in general unknown, let us briefly make contact
with this method as we will repeatedly refer to it in our subsequent
discussion.

To start, we note that the existence of a map $\eta $ with the
aforementioned properties implies that $H$ and$~H^{\ast }$ have real
spectrum. However, the converse of this statement is in general \emph{not}
true. Suppose we are given a finite-dimensional non-Hermitian Hamiltonian
with $H,$ such that $H$ and $H^{\ast }$ have purely real spectra, $\limfunc{%
Spec}H=\limfunc{Spec}H^{\ast }\subset \mathbb{R}.$ Then there exist two
corresponding sets of eigenvectors $\{\phi _{\lambda }\}_{\lambda \in 
\limfunc{Spec}H}$ of $H$ and $\{\psi _{\lambda }\}_{\lambda \in \limfunc{Spec%
}H}$ of $H^{\ast }$ such that%
\begin{equation}
H\phi _{\lambda }=\lambda \phi _{\lambda }\qquad \text{and\qquad }H^{\ast
}\psi _{\lambda }=\lambda \psi _{\lambda }\ .  \label{eigen}
\end{equation}%
Making the \emph{additional} assumption that the eigenvectors can be
normalized such that the relations%
\begin{equation}
\langle \phi _{\lambda },\psi _{\lambda ^{\prime }}\rangle =\delta _{\lambda
,\lambda ^{\prime }}\qquad \text{and}\qquad \sum_{\lambda \in \limfunc{Spec}%
H}|\phi _{\lambda }\rangle \langle \psi _{\lambda }|~=\mathbf{1}
\label{biortho}
\end{equation}%
both hold, one can then define $\eta $ to be the following sum over
projectors%
\begin{equation}
\eta =\sum_{\lambda \in \limfunc{Spec}H}|\psi _{\lambda }\rangle \langle
\psi _{\lambda }|~=\left( \sum_{\lambda \in \limfunc{Spec}H}|\phi _{\lambda
}\rangle \langle \phi _{\lambda }|\right) ^{-1}\ .  \label{dynam eta}
\end{equation}%
The above map $\eta $ satisfies all necessary requirements by construction.
We stress, however, that the crucial assumption concerning the existence of
a bi-orthonormal system of eigenvectors satisfying (\ref{biortho}) is not
always satisfied.

In particular, the spin-chain Hamiltonian (\ref{H}) provides a
counterexample. Namely, if we choose $q$ to be a primitive root of unity,
then one can verify that the Hamiltonian possesses non-trivial Jordan
blocks. As a result there always exist eigenvalues $\lambda \in \limfunc{Spec%
}H$ for which the associated eigenvectors $\phi _{\lambda },\psi _{\lambda }$
are orthogonal,%
\begin{equation}
\langle \phi _{\lambda },\psi _{\lambda }\rangle =0\;.  \label{selfortho}
\end{equation}%
The connection between such states and non-trivial Jordan blocks has been
discussed for example in \cite{SAC06,GRS07}. The occurrence of such states
in the present context can be understood in terms of representation theory 
\cite{PS90}. Due to (\ref{selfortho}), the assumption (\ref{biortho}) is
violated and one only finds a positive semi-definite matrix $\eta ,$ $\eta
\geq 0,$ when $q$ is a root of unity. The solution to this problem of
defining a Hermitian version of the quantum group invariant XXZ Hamiltonian
at roots of unity is to remove the aforementioned subset of states from the
Hilbert space, making $\eta $ positive definite, $\eta >0$, and,
consequently, the Hamiltonian diagonalizable. This procedure is known as
"quantum group reduction at roots of unity". This reduction procedure has
been developed previously in the literature on integrable systems \cite{RS90}%
. The new aspect of our discussion here is its relation to the ideas of
quasi-Hermiticity. Developing this connection will put us into the position
to present in the last part of this paper a novel algebraic formulation of
the Hilbert space structure which makes (\ref{H}) Hermitian.

\subsection{The path basis}

In order to reduce the state space we first introduce a different set of
basis vectors. The new set of basis vectors which we are going to construct
is dictated by the quantum group invariance of the Hamiltonian and has the
advantage that the set of "problematic" states (\ref{selfortho}) can be
easily identified. For the moment, we keep $q=\exp (i\pi /r)$ generic, but
shall specialize below to integer values $r\geq 3$.

The new basis states will be obtained by successively decomposing tensor
products of $U_{q}(sl_{2})$ representations into the finite-dimensional
irreducible representations $\pi _{j}:U_{q}(sl_{2})\rightarrow \limfunc{End}%
\mathbb{C}^{2j+1}$ with $j\in \frac{1}{2}\mathbb{N}$. Up to isomorphism the
latter are given by%
\begin{eqnarray}
\pi _{j}(s^{\pm })\left\vert j,m\right\rangle &=&\sqrt{[j\mp m]_{q}[j\pm
m+1]_{q}}\left\vert j,m\pm 1\right\rangle ,  \notag \\
\pi _{j}(q^{s^{z}})\left\vert j,m\right\rangle &=&q^{m}\left\vert
j,m\right\rangle ,\;\;m=-j,-j+1,...,j-1,j\;.  \label{spin j rep}
\end{eqnarray}%
In analogy with the terminology used for $sl_{2}$, the half-integer $j$
labeling the representation is referred to as "spin". Following \cite%
{RS90,JK94} we now introduce a "path basis" using the $q$-deformed
Clebsch-Gordan (CG) coefficients defined implicitly via the embedding%
\begin{equation}
\imath _{12}:\pi _{J}\hookrightarrow \pi _{j_{1}}\otimes \pi _{j_{2}},\quad
\left\vert J,M\right\rangle \hookrightarrow \sum_{m_{1}+m_{2}=M}\left\vert 
\begin{array}{ccc}
j_{1} & j_{2} & J \\ 
m_{1} & m_{2} & M%
\end{array}%
\right\vert _{q}\;\left\vert j_{1},m_{1}\right\rangle \otimes \left\vert
j_{2},m_{2}\right\rangle \ .  \label{inc}
\end{equation}%
The relevant CG coefficients for the spin $1/2$ chain are computed via the
action of the coproduct (see Appendix C),%
\begin{equation}
\left\vert 
\begin{array}{ccc}
j & \frac{1}{2} & j+\frac{1}{2} \\ 
m & \alpha & m+\alpha%
\end{array}%
\right\vert _{q}=q^{-\alpha j+\frac{m}{2}}\left( \frac{[j+2\alpha m+1]}{%
[2j+1]}\right) ^{\frac{1}{2}}\   \label{cgc1}
\end{equation}%
and%
\begin{equation}
\left\vert 
\begin{array}{ccc}
j & \frac{1}{2} & j-\frac{1}{2} \\ 
m & \alpha & m+\alpha%
\end{array}%
\right\vert _{q}=2\alpha q^{\alpha (j+1)+\frac{m}{2}}\left( \frac{[j-2\alpha
m]}{[2j+1]}\right) ^{\frac{1}{2}}\;.  \label{cgc2}
\end{equation}%
Now let $\boldsymbol{j}=(j_{0},j_{1},j_{2},...,j_{N})$ be a path on the $%
sl_{2}$-Bratelli diagram, i.e. the set of sequences specified as follows%
\begin{equation}
\Gamma =\{\boldsymbol{j}=(j_{0},j_{1},j_{2}...,j_{N})~|~j_{0}=0,\;j_{k}\geq
0,\;j_{k+1}=j_{k}\pm 1/2\}\ .  \label{bratelli}
\end{equation}%
Here we have followed the common convention to root the paths at $j_{0}=0$
which forces $j_{1}=1/2$. Next, we define the vectors%
\begin{equation}
\left\vert \boldsymbol{j},m\right\rangle =\sum_{|\mathbf{\alpha }%
|=m}\left\vert \mathbf{\alpha }\right\rangle \left\langle \mathbf{\alpha }|%
\boldsymbol{j},m\right\rangle ,\qquad m=-j_{N},-j_{N}+1,...,0,...,j_{N}
\label{path state}
\end{equation}%
with%
\begin{equation}
\left\langle \mathbf{\alpha }|\boldsymbol{j},m\right\rangle
=\prod_{k=1}^{N-1}\left\vert 
\begin{array}{ccc}
j_{k} & \frac{1}{2} & j_{k+1} \\ 
\sum_{i\leq k}\alpha _{i} & \alpha _{k+1} & \sum_{i\leq k+1}\alpha _{i}%
\end{array}%
\right\vert _{q},\qquad m=|\mathbf{\alpha }|:=\sum_{k=1}^{N}\alpha
_{k}=S^{z}\ .  \label{path matrix}
\end{equation}%
As long as $q$ is not a root of unity the above basis is well-defined. If $r$
is an integer $\geq 3$, however, one has to constrain the set of allowed
paths to the restricted Bratelli diagram 
\begin{equation}
\Gamma ^{(r)}:=\{\boldsymbol{j}\in \Gamma ~|~2j_{k}+1<r,\;k=1,...,N\}\;.
\label{RSOS}
\end{equation}%
This restriction ensures that no singularities or cancellations occur in the
CG coefficients due to the factors $[2j_{k}+1]_{q}=[r]_{q}=0$. It is
precisely the reduction of the state space by the constraint (\ref{RSOS})
which removes the states (\ref{selfortho}) mentioned above in the context of
quasi-Hermiticity. As we will see below, we can then explicitly construct a
positive map $\eta $ on the reduced state space.

\subsection{Action of the Temperley-Lieb algebra in the path basis}

The introduction of the path basis is not only motivated by aspects of
quasi-Hermiticity but is rather natural from an algebraic point of view. It
allows us to factor out the quantum group action which commutes with the
Hamiltonian and spans its degenerate eigenspaces. In fact, given a fixed
path $\boldsymbol{j}$ on the Bratelli-diagram the action of the quantum
group $U_{q}(sl_{2})$ will not change this path but only modify the
corresponding "magnetic quantum number" $m$ in the associated path state (%
\ref{path state}), i.e. each path $\boldsymbol{j}%
=(j_{0}=0,j_{1}=1/2,j_{2},...,j_{N})\in \Gamma $ corresponds to an
irreducible quantum group module of spin $j_{N}$ given by (\ref{spin j rep}).

The Hamiltonian on the other hand can only mix paths with the same end point 
$j_{N}$ and leaves $m$ unchanged due to (\ref{QGinv}). The same holds true
for the generators of the Temperley-Lieb algebra. Both assertions follow
immediately from the quantum version of Schur-Weyl duality (\ref{QBFS
duality}). They can be explicitly verified by computing the action of the
Temperley-Lieb generators (\ref{TLrep}) in the path basis from (\ref{cgc1}),
(\ref{cgc2}) and (\ref{path state}); see Appendix C for the relevant
identities. One finds for $k=2,...,N-1$ the following result,%
\begin{equation}
E_{k}\left\vert \boldsymbol{j},m\right\rangle =\delta _{j_{k-1},j_{k+1}}%
\hspace*{-6mm}\sum_{j^{\prime }=j_{k-1}\pm 1/2}\hspace*{-6mm}\frac{\sqrt{%
[2j_{k}+1]_{q}[2j^{\prime }+1]_{q}}}{(-)^{j_{k}-j^{\prime
}+1}[2j_{k-1}+1]_{q}}~\left\vert j_{0},j_{1},...j_{k-1},j^{\prime
},j_{k+1},...j_{N},m\right\rangle \ ,  \label{pathTL}
\end{equation}%
where $\left\vert \boldsymbol{j},m\right\rangle =0$ if $j_{i}<0$ for some $i$%
. Note, that for $k=1$ the formula (\ref{pathTL}) therefore simplifies to%
\begin{equation}
E_{1}\left\vert \boldsymbol{j},m\right\rangle =-\delta
_{0,j_{2}}[2]_{q}\left\vert \boldsymbol{j},m\right\rangle ,
\end{equation}%
where we have used the expressions%
\begin{equation}
\left\vert 
\begin{array}{ccc}
\frac{1}{2} & \frac{1}{2} & 1 \\ 
\alpha _{1} & \alpha _{2} & m%
\end{array}%
\right\vert _{q}=q^{\frac{\alpha _{1}-\alpha _{2}}{2}}\left( \frac{[\frac{3}{%
2}+2\alpha _{1}\alpha _{2}]}{[2]}\right) ^{\frac{1}{2}},\qquad \left\vert 
\begin{array}{ccc}
\frac{1}{2} & \frac{1}{2} & 0 \\ 
\alpha _{1} & \alpha _{2} & 0%
\end{array}%
\right\vert _{q}=-2\alpha _{1}q^{-\alpha _{1}}/[2]^{\frac{1}{2}}\ .
\end{equation}%
Thus, $E_{1}$ is a diagonal matrix in the path basis. From the expression (%
\ref{pathTL}), one can now directly read off the previously stated
properties of the action of the Temperley-Lieb algebra and the Hamiltonian.
As far as the Temperley-Lieb action is concerned, one may disregard the
magnetic quantum number $m$ in (\ref{path state}) and concentrate on the
subspaces 
\begin{equation}
\Gamma _{j}:=\{\boldsymbol{j}\in \Gamma ~|~\boldsymbol{j}%
=(0,1/2,j_{2},...,j_{N-1},j_{N}=j)\}  \label{Gammaj}
\end{equation}%
with fixed end point $j$ which according to (\ref{pathTL}) are invariant. In
an analogous fashion we define for $q=\exp (i\pi /r)$ with $r$ integer the
set%
\begin{equation}
\Gamma _{j}^{(r)}:=\Gamma ^{(r)}\cap ~\Gamma _{j}\ .  \label{Gammarj}
\end{equation}%
The representations $\Gamma _{j}$ and $\Gamma _{j}^{(r)}$ can respectively
be shown to be equivalent to the finite-dimensional irreducible
representations of the Temperley-Lieb algebra at generic $q$ and at roots of
unity $q^{r}=-1$ given for example in \cite{Jo83,We88,We90,PPM91} and
references therein. We discuss an alternative, graphical description of the
irreducible representations of $TL_{N}(q)$ in Appendix A.\medskip

\begin{center}
\begin{tabular}{||c||c|c|c|c|c|c|c|c|c|c|c|}
\hline
$N~\backslash ~j$ & 0 & $\frac{1}{2}$ & 1 & $\frac{3}{2}$ & 2 & $\frac{5}{2}$
& 3 & $\frac{7}{2}$ & 4 & $\frac{9}{2}$ & 5 \\ \hline\hline
0 & 1 &  &  &  &  &  &  &  &  &  &  \\ \cline{1-1}\cline{1-3}\cline{2-3}
1 &  & 1 &  &  &  &  &  &  &  &  &  \\ \cline{1-4}\cline{4-4}
2 & 1 &  & 1 &  &  &  &  &  &  &  &  \\ \cline{1-5}
3 &  & 2 &  & 1 &  &  &  &  &  &  &  \\ \cline{1-6}
4 & 2 &  & 3 &  & 1 &  &  &  &  &  &  \\ \cline{1-7}
5 &  & 5 &  & 4 &  & 1 &  &  &  &  &  \\ \cline{1-8}
6 & 5 &  & 9 &  & 5 &  & 1 &  &  &  &  \\ \cline{1-9}
7 &  & 14 &  & 14 &  & 6 &  & 1 &  &  &  \\ \cline{1-10}
8 & 14 &  & 28 &  & 20 &  & 7 &  & 1 &  &  \\ \cline{1-11}
9 &  & 42 &  & 48 &  & 27 &  & 8 &  & 1 &  \\ \hline
10 & 42 &  & 90 &  & 75 &  & 35 &  & 9 &  & 1 \\ \hline
\end{tabular}%
~%
\begin{tabular}{||c||c|c|c|c|}
\hline
$N~\backslash ~j$ & 0 & $\frac{1}{2}$ & 1 & $\frac{3}{2}$ \\ \hline\hline
0 & 1 &  &  &  \\ \cline{1-1}\cline{1-3}\cline{2-3}
1 &  & 1 &  &  \\ \cline{1-4}\cline{4-4}
2 & 1 &  & 1 &  \\ \hline
3 &  & 2 &  & 1 \\ \hline
4 & 2 &  & 3 &  \\ \hline
5 &  & 5 &  & 3 \\ \hline
6 & 5 &  & 8 &  \\ \hline
7 &  & 13 &  & 8 \\ \hline
8 & 13 &  & 21 &  \\ \hline
9 &  & 34 &  & 21 \\ \hline
10 & 34 &  & 55 &  \\ \hline
\end{tabular}%
\medskip \medskip

Table 2. Dimensions $\dim \Gamma _{j}$ and $\dim \Gamma _{j}^{(5)}$ for
different chains of length $N$.\bigskip
\end{center}

\noindent When decomposing the Hilbert space $V^{\otimes N}$ into the
irreducible representations $\pi _{j}$ of the quantum group for generic $q$,
the following multiplicities $\mu _{j}$ occur which coincide with the
dimension of the path space $\Gamma _{j}$,%
\begin{equation}
\dim \Gamma _{j}=\mu _{j}:=\dbinom{N}{N/2-j}-\dbinom{N}{N/2+j+1}\ .
\label{dimGamj}
\end{equation}%
For $q=\exp (i\pi /r)$ with $r$ integer~$\geq 3$, the dimension of the
irreducible representations $\Gamma _{j}^{(r)}$ are obtained via the formula,%
\begin{equation}
\dim \Gamma _{j}^{(r)}=\sum_{k=-\infty }^{\infty }\mu _{j+rk},
\label{dimGamrj}
\end{equation}%
where the sum always turns out to contain only a finite number of
non-vanishing terms, since $\dbinom{m}{n}=0$ for $n<0$ or $n>m$. An
illustration of the multiplicities is given in Table 2.

\subsection{Path basis construction of $\protect\eta $ at roots of unity}

We are now in the position to construct the map $\eta $ described in the
introduction which will allow us to establish quasi-Hermiticity for the
Hamiltonian (\ref{H}). As a preparatory step, we first introduce the path
basis with respect to the Hermitian adjoint Hamiltonian $H^{\ast }$. Using
the time reversal operator (i.e. complex conjugation with respect to the
spin basis (\ref{spin-basis})) we define the conjugate path basis%
\begin{equation}
\left\vert \boldsymbol{j},m\right\rangle _{T}:=T\left\vert \boldsymbol{j}%
,m\right\rangle =\sum_{|\mathbf{\alpha }|=m}\left\vert \mathbf{\alpha }%
\right\rangle \overline{\left\langle \mathbf{\alpha }|\boldsymbol{j}%
,m\right\rangle },\qquad \boldsymbol{j}\in \Gamma
,\;m=-j_{N},-j_{N}+1,...,j_{N}\ .  \label{Tpath state}
\end{equation}%
Using the unitarity relation%
\begin{equation}
\sum_{m,\alpha }\left\vert 
\begin{array}{ccc}
j & \frac{1}{2} & j^{\prime } \\ 
m & \alpha & m^{\prime }%
\end{array}%
\right\vert _{q}\left\vert 
\begin{array}{ccc}
j & \frac{1}{2} & j^{\prime \prime } \\ 
m & \alpha & m^{\prime \prime }%
\end{array}%
\right\vert _{q}=\delta _{j^{\prime },j^{\prime \prime }}\delta _{m^{\prime
},m^{\prime \prime }}  \label{unitarity}
\end{equation}%
we have the following identities%
\begin{equation}
_{T}\!\left\langle \boldsymbol{j},m|\boldsymbol{j}^{\prime },m^{\prime
}\right\rangle =\delta _{m,m^{\prime }}\prod_{k}\delta _{j_{k},j_{k}^{\prime
}}\text{\qquad and\qquad }\mathbf{1}=\sum_{\boldsymbol{j},m}\left\vert 
\boldsymbol{j},m\right\rangle ~_{T}\!\left\langle \boldsymbol{j}%
,m\right\vert \ .  \label{completeness}
\end{equation}%
The above relations hold for all values of $q$ on the unit circle. In
particular, the completeness relation (the second identity in (\ref%
{completeness})) continues to be true when $q$ is a root of unity, $q=\exp
(i\pi /r)$ with $r$ integer~$\geq 3$, and the path space is reduced to (\ref%
{RSOS}).$\boldsymbol{\smallskip }$ \textbf{For the remainder of this section
we shall assume that we are at such a special root of unity value.}$%
\boldsymbol{\smallskip }$ \medskip

\noindent \underline{\textsc{Definition 4.1}}\textsc{\ [}$\eta $\textsc{\ at
roots of 1].} \emph{Let }$q=e^{i\pi /r}$\emph{\ with }$r>2$\emph{\ and
integer. Then we define the quasi-Hermiticity operator }$\eta $\emph{\ to be}%
\begin{equation}
\eta =\sum_{\boldsymbol{j},m}\left\vert \boldsymbol{j},m\right\rangle
_{T}~_{^{{}}T}\!\left\langle \boldsymbol{j},m\right\vert ~,  \label{eta}
\end{equation}%
\emph{where the sum ranges over all restricted paths }$\boldsymbol{j}\in
\Gamma ^{(r)}$\emph{\ defined in (\ref{RSOS}) and }$m$\emph{\ over the
values given in (\ref{Tpath state}).}$\boldsymbol{\smallskip }$

By definition, $\eta $ is positive definite. Its expression in terms of
projectors onto path states is reminiscent of similar expressions in terms
of bi-orthonormal systems of eigenvectors as they can be found in the
literature on $PT$-symmetry and quasi-Hermiticity; compare with (\ref{eigen}%
), (\ref{biortho}) and (\ref{dynam eta}). We wish to stress, however, that
the path states are \emph{not }eigenvectors of the Hamiltonian and that they
are explicitly given through the CG coefficients (\ref{cgc1}), (\ref{cgc2})
and (\ref{Tpath state}). Thus, we have a \emph{non-perturbative, exact and
explicit} expression for the quasi-Hermiticity operator $\eta $ without
having to rely on the construction of the eigenvectors of the Hamiltonian.
To emphasize this point even further, we explicitly state the matrix
elements of $\eta $ in the local spin-basis, 
\begin{eqnarray}
\left\langle \mathbf{\alpha }|\eta |\mathbf{\beta }\right\rangle &=&\sum_{%
\boldsymbol{j},m}\overline{\left\langle \mathbf{\alpha }|\boldsymbol{j}%
,m\right\rangle }\left\langle \mathbf{\beta }|\boldsymbol{j},m\right\rangle 
\notag \\
&=&\sum_{\boldsymbol{j},m}\prod_{k=1}^{M-1}\overline{\left\vert 
\begin{array}{ccc}
j_{k} & \frac{1}{2} & j_{k+1} \\ 
m_{k} & \alpha _{k+1} & m_{k+1}%
\end{array}%
\right\vert }_{q}\left\vert 
\begin{array}{ccc}
j_{k} & \frac{1}{2} & j_{k+1} \\ 
m_{k}^{\prime } & \beta _{k+1} & m_{k+1}^{\prime }%
\end{array}%
\right\vert _{q}\ .  \label{etamatrix}
\end{eqnarray}%
It remains to show that the map $\eta $ defined by (\ref{eta}) intertwines
the Hamiltonian (\ref{H}) with its Hermitian adjoint. In fact we find that
it obeys more stringent conditions.\smallskip

\noindent \underline{\textsc{Proposition 4.1}}\textsc{.} \emph{As before let 
}$q=\exp (i\pi /r)$\emph{\ with }$r$ \emph{an integer~}$>2$\emph{. Then the
map }$\eta $\emph{\ defined in (\ref{eta}) satisfies the identity}%
\begin{equation}
\eta E_{k}=E_{k}^{\ast }\eta \ .  \label{etaTL}
\end{equation}%
\emph{Here }$\{E_{k}\}_{k=1}^{N-1}$\emph{\ are the Temperley-Lieb generators
(\ref{TLdef}) in the representation (\ref{pathTL}) and under the restriction
(\ref{RSOS}). In addition, the following equalities hold for the quantum
group generators}%
\begin{equation}
\eta S^{\pm }=(S^{\mp })^{\ast }\eta =S_{\text{op}}^{\pm }\eta \qquad \text{%
\emph{and}}\qquad \lbrack \eta ,S^{z}]=0\ .  \label{etainter}
\end{equation}%
\emph{Again these identities are valid only after the reduction of the state
space according to (\ref{RSOS}) has been imposed.}\smallskip 

\noindent \textsc{Proof.} In order to derive the first identity we note that
the matrix elements of the Temperley-Lieb generators in (\ref{pathTL}) are
real numbers as long as all $q$-integers appearing in the above equation (%
\ref{pathTL}) are positive. This is guaranteed by the choice of $q$ and the
restriction of the paths to the Bratelli diagram $\Gamma ^{(r)}$ in (\ref%
{RSOS}). This explains the first equality in the sequence 
\begin{equation}
\eta E_{k}\left\vert \boldsymbol{j},m\right\rangle \ =TE_{k}\left\vert 
\boldsymbol{j},m\right\rangle =E_{k}^{\ast }T\left\vert \boldsymbol{j}%
,m\right\rangle =E_{k}^{\ast }\eta \left\vert \boldsymbol{j},m\right\rangle .
\end{equation}%
The second equality follows from the observation that the Temperley-Lieb
generators are represented as symmetric matrices. The final equality arises
simply from the definition (\ref{eta}). Together with the fact that the path
states form a basis this proves (\ref{etaTL}). In a similar fashion one
proves the second identity for the quantum group generators.\ $\blacksquare $%
\smallskip

We then have as a trivial consequence of the above proposition the desired
property%
\begin{equation}
\eta H=H^{\ast }\eta ,\qquad H=\sum_{k=1}^{N-1}E_{k}\ .
\end{equation}%
Thus, we have also demonstrated that $H$ is indeed quasi-Hermitian and that (%
\ref{eta}) defines the physically relevant inner product (\ref{etaprod})
with respect to which $H$ is Hermitian. This is particularly important for
the calculation of physically relevant quantities such as correlation
functions where one considers matrix elements of local operators \cite%
{HMRS93}. While the expression (\ref{etamatrix}) is especially convenient
for numerical computations, an alternative formulation in terms of the
relevant algebras would be desirable in order to apply more powerful
mathematical techniques. In light of (\ref{etaTL}) and (\ref{etainter}), we
can immediately conclude that $\eta $ cannot be expressed in terms of the
quantum group or Temperley-Lieb algebra; see (\ref{QBFS duality}). We
therefore turn our attention to the $C$-operator.

\section{The $C$-operator}

In the context of $PT$-symmetry, Bender and collaborators introduced the $C$%
-operator in order to construct a well-defined inner product with respect to
which the Hamiltonian becomes Hermitian; for references see \cite{Bender07}.
It is now understood that this approach is a special case of
quasi-Hermiticity.\smallskip

\noindent \underline{\textsc{Definition 5.1}}\textsc{\ [the C-operator].} 
\emph{Given a positive, Hermitian and invertible map }$\eta :\mathfrak{H}%
\rightarrow \mathfrak{H}$ \emph{with} $\eta H=H^{\ast }\eta ,$ \emph{the }$C$%
\emph{-operator is defined to be the linear map} 
\begin{equation}
C=P\eta ,  \label{etaC}
\end{equation}%
\emph{where }$P$\emph{\ is the previously defined parity operator which is
assumed to obey (\ref{PHP})}.\smallskip

An immediate consequence of the above definition is the relation,%
\begin{equation}
\lbrack H,C]=0\ .  \label{HC}
\end{equation}%
It should be clear that the choice of the parity operator in (\ref{etaC}) is
only dictated by convenience. For instance, in the present case - due to (%
\ref{RHR}) - we might equally well choose the spin-reversal operator $R$
instead of $P$ which leads to a second, alternative definition.\smallskip

\noindent \underline{\textsc{Definition 5.2}}\textsc{\ [the C}$^{\prime }$%
\textsc{-operator].} \emph{For given }$\eta $ \emph{as in the previous
definition, we set}%
\begin{equation}
C^{\prime }=R\eta \;.  \label{etaC'}
\end{equation}%
\smallskip\ Again we have that $[H,C^{\prime }]=0$. In contrast, the choice
of the time reversal operator $T$ for the definition of $C$ would not be on
the same footing, because $T$ is antilinear. In the literature on $PT$%
-symmetry, one usually finds the additional requirement that $C$ is an
involution, i.e.%
\begin{equation}
C^{2}=1  \label{C2}
\end{equation}%
or equivalently that 
\begin{equation}
P\eta P=\eta ^{-1}~.  \label{PetaP}
\end{equation}%
In many examples this turns out to be true. In particular, if (\ref{PHP})
holds and a bi-orthonormal eigensystem (\ref{eigen}), (\ref{biortho}) can be
chosen such that $P\phi _{\lambda }=\psi _{\lambda }$, the properties (\ref%
{C2}) and (\ref{PetaP}) immediately follow from (\ref{dynam eta}).
Nevertheless, (\ref{C2}) is not a fundamental property necessary to ensure
that the inner product (\ref{etaprod}) is well defined. For this reason we
have not included this property in the definition of $C$. However, we will
show below that (\ref{C2}) does indeed hold true in the present construction
and that the analogous relation is satisfied by $C^{\prime }$ as well.

Our motivation to consider the two aforementioned $C$-operators becomes
clear when looking at their commutation relations. As we well discuss below,
the operator $C$ commutes with the quantum group action while $C^{\prime }$
commutes with the Temperley-Lieb action. We will use these commutation
relations to describe the properties of $C,C^{\prime }$ and give algebraic
construction for both.

\subsection{Properties and identities of the $C$-operators}

We start the discussion with the operator $C^{\prime }$ as its action in the
path basis is considerably simpler than the action of $C$. In a second step,
we shall then use these results to find an elegant algebraic expression for
the operator $C$ in terms of the Hecke algebra.\smallskip

\noindent \underline{\textsc{Theorem 5.1}}\textsc{.} \emph{Let }$q=\exp
(i\pi /r)$ \emph{with }$r$\emph{\ integer}$\geq 3$\emph{\ and }$\eta $ \emph{%
be the previously defined sum over projectors in the path basis (\ref{eta}).
Set }$C^{\prime }=R\eta $\emph{\ with }$R$ \emph{the spin-reversal operator.
Then }%
\begin{equation}
C^{\prime }\left\vert \boldsymbol{j},m\right\rangle =(-)^{\frac{N}{2}%
-j_{N}}\left\vert \boldsymbol{j},-m\right\rangle ,  \label{C'action}
\end{equation}%
\emph{where }$j_{N}$\emph{\ is the endpoint of the path }$\boldsymbol{j}$%
\emph{. Thus, we have in particular that }$C^{\prime 2}=1$ \emph{or
equivalently}%
\begin{equation}
R\eta R=\eta ^{-1}\ .  \label{RetaRinv}
\end{equation}%
\emph{According to (\ref{spin j rep}) and (\ref{C'action}) the operator }$%
C^{\prime }$\emph{\ can be expressed on }$\Gamma _{j}^{(r)}$\emph{\ in terms
of the quantum group generators as}%
\begin{equation}
C^{\prime }|_{\Gamma _{j}^{(r)}}=(-)^{\frac{N}{2}-j}\sum_{m\in \frac{1}{2}%
\mathbb{N}}\frac{[j-m]_{q}!}{[j+m]_{q}!}~\frac{(S^{-})^{2m}\delta
_{S_{z},m}+(S^{+})^{2m}\delta _{S_{z},-m}}{2^{\delta _{0,m}}}.  \label{C'S}
\end{equation}%
%
%
%
%
%
%
%
%
%
%
%
%
%\emph{where }$[0]_{q}!\equiv 1$. 
\emph{This gives an expression for} $C^{\prime }$\emph{which is independent
of the path basis.}\medskip

\noindent \textsc{Proof.} First we note from (\ref{etaTL}) and (\ref{RTLR})
that $C^{\prime }=R\eta $ commutes with the Temperley-Lieb algebra and hence
we can conclude from (\ref{QBFS duality}) that $C^{\prime }\in \mathcal{U}$.
Since each subspace $\Gamma _{j}^{(r)}$ of restricted paths with the same
endpoint $j_{N}=j$ forms an irreducible, faithful representation of the
Temperley-Lieb algebra, it suffices to compute the action of $C^{\prime }$
on $\left\vert \boldsymbol{j}^{\prime },m\right\rangle $ for some special
path $\boldsymbol{j}^{\prime }\in \Gamma _{j}^{(r)}$ in order to infer its
action on any path state $\left\vert \boldsymbol{j},m\right\rangle ,\ 
\boldsymbol{j}\in \Gamma _{j}^{(r)}$. In other words, $C^{\prime }$ can at
most change the magnetic quantum number $m$ but not the actual path $%
\boldsymbol{j}$ in a path state $\left\vert \boldsymbol{j},m\right\rangle $.

For given $j_{N}$ let us pick the path $\boldsymbol{j}^{\prime }$ on the
Bratelli diagram which alternates a maximal number of times between $j=0$
and $j=1/2$ before increasing monotonically to $j=j_{N}$ (for example, if $%
N=6$ and $j_{6}=2$, we would choose $\boldsymbol{j}^{\prime
}=(0,1/2,0,1/2,1,3/2,2)$). Then the path states $\left\vert \boldsymbol{j}%
^{\prime },\pm j_{N}\right\rangle $ consist of the following linear
combination of vectors in the spin basis,%
\begin{equation}
\left\vert \boldsymbol{j}^{\prime },\pm j_{N}\right\rangle =\sum_{\alpha
_{k}}~|\alpha _{1},-\alpha _{1},...,\alpha _{N-2j_{N}},-\alpha _{N-2j_{N}},%
\underset{2j_{N}}{\underbrace{\pm \tfrac{1}{2},...,\pm \tfrac{1}{2}}}\rangle
\prod_{k=1}^{\frac{N}{2}-j_{N}}\left( -\frac{2\alpha _{k}q^{-\alpha _{k}}}{%
[2]_{q}^{1/2}}\right) \ .  \label{jprime}
\end{equation}%
Here we have used the following identities for the Clebsch-Gordan
coefficients in the path basis expansion (\ref{path state}),%
\begin{equation}
\prod_{k=1}^{\frac{N}{2}-j_{N}}\left\vert 
\begin{array}{ccc}
\frac{1}{2} & \frac{1}{2} & 0 \\ 
\alpha _{k} & -\alpha _{k} & 0%
\end{array}%
\right\vert _{q}\left\vert 
\begin{array}{ccc}
0 & \frac{1}{2} & \frac{1}{2} \\ 
0 & \alpha _{k+1} & \alpha _{k+1}%
\end{array}%
\right\vert _{q}=\prod_{k=1}^{\frac{N}{2}-j_{N}}\left( -\frac{2\alpha
_{k}q^{-\alpha _{k}}}{[2]_{q}^{1/2}}\right)
\end{equation}%
and%
\begin{equation}
\prod_{k=0}^{2j_{N}-1}\left\vert 
\begin{array}{ccc}
\frac{k}{2} & \frac{1}{2} & \frac{k+1}{2} \\ 
\pm \frac{k}{2} & \pm \frac{1}{2} & \pm \frac{k+1}{2}%
\end{array}%
\right\vert _{q}=1\ .
\end{equation}%
From the above expressions one now easily verifies that%
\begin{equation}
C^{\prime }\left\vert \boldsymbol{j}^{\prime },j_{N}\right\rangle
=R\left\vert \boldsymbol{j}^{\prime },j_{N}\right\rangle _{T}=(-)^{\frac{N}{2%
}-j_{N}}\left\vert \boldsymbol{j}^{\prime },-j_{N}\right\rangle \ .
\end{equation}%
But (\ref{etainter}) and (\ref{RSR}) now imply 
\begin{eqnarray}
C^{\prime }\left\vert \boldsymbol{j}^{\prime },m\right\rangle &=&\mathcal{N}%
_{m}C^{\prime }(S^{-})^{j_{N}-m}\left\vert \boldsymbol{j}^{\prime
},j_{N}\right\rangle =\mathcal{N}_{m}(S^{+})^{j_{N}-m}C^{\prime }\left\vert 
\boldsymbol{j}^{\prime },j_{N}\right\rangle  \notag \\
&=&(-)^{\frac{N}{2}-j_{N}}\mathcal{N}_{m}(S^{+})^{j_{N}-m}\left\vert 
\boldsymbol{j}^{\prime },-j_{N}\right\rangle =(-)^{\frac{N}{2}%
-j_{N}}\left\vert \boldsymbol{j}^{\prime },-m\right\rangle \ .
\end{eqnarray}%
Here $\mathcal{N}_{m}=\sqrt{[j_{N}+m]_{q}!/[2j_{N}]_{q}![j_{N}-m]_{q}!}$ is
some unimportant normalization constant; see (\ref{spin j rep}). The
expression (\ref{C'S}) now also follows from this result. Note that (\ref%
{C'S}) is consistent with the quantum Schur-Weyl duality (\ref{QBFS duality}%
), $C^{\prime }\in \mathcal{U}$, since the Kronecker $\delta $-functions can
be rewritten in terms of the quantum group generators $q^{\pm S^{z}},$%
\begin{equation}
\delta _{S_{z},\pm m}=r^{-1}\sum_{k=1}^{r}q^{4(m\mp S^{z})k}\ .
\end{equation}%
$\blacksquare $

From the above theorem we infer that the action of $C^{\prime }$ is
surprisingly simple; a result which is not obvious given the expression (\ref%
{eta}) involving a sum of projectors. However, we stress that the simple
expression (\ref{C'action}) is special to the path basis construction.

We now state the analogous result for the $C$-operator and give its explicit
algebraic form.\smallskip

\noindent \underline{\textsc{Theorem 5.2}}\textsc{.} \emph{Let }$\eta $\emph{%
\ be defined as before and set }$C=P\eta $\emph{. Then we have that} 
\begin{equation}
\lbrack C,C^{\prime }]=0\qquad \text{\emph{and}\qquad }C^{2}=1\;.
\label{CC'0}
\end{equation}%
\emph{Furthermore, upon restriction to the invariant subspaces }$\Gamma
_{j}^{(r)}$\ \emph{the following operator identity holds:}%
\begin{equation}
C|_{\Gamma _{j}^{(r)}}=\chi _{j}\mathcal{B},\qquad \chi _{j}\in \mathbb{C}\ .
\label{Cbeta}
\end{equation}%
\emph{Here }$\mathcal{B}$ \emph{denotes the image of the following special
braid }$\beta $ \emph{under the representation (\ref{Hecke spin})
respectively (\ref{pathTL}) induced via (\ref{hom}),}%
\begin{equation}
\beta =\beta _{1}\beta _{2}\cdots \beta _{N-1},\qquad \beta
_{n}=b_{n}b_{n-1}\cdots b_{1}\ .  \label{beta}
\end{equation}%
\smallskip

\noindent \textsc{Proof.} From the transformations (\ref{etainter}) and (\ref%
{PSP}) we infer that the $C$-operator is quantum group invariant, $[C,%
\mathcal{U}]=0$. That is, according to the quantum analogue of Schur-Weyl
duality (\ref{QBFS duality}) we must have $C\in \mathcal{H}$. We already saw
in the proof of the previous theorem that $[C^{\prime },\mathcal{H}]=0$,
whence $C^{\prime }\in \mathcal{U}$. Hence, the first assertion, $%
[C,C^{\prime }]=0$, trivially follows from (\ref{QBFS duality}). But due to
the fact that $[P,R]=0,$ the commutation of the two $C$-operators is
equivalent to $C^{2}=C^{\prime 2}=1$; see Theorem 5.1.

The last assertion (\ref{Cbeta}) is now deduced by first noting that (\ref%
{PTLP}) and (\ref{etaTL}) imply that $C$ obeys 
\begin{equation}
CE_{k}=P\eta E_{k}=E_{N-k}C\ .
\end{equation}%
In other words, $C$ corresponds to parity reversal within the Temperley-Lieb
algebra, i.e. it implements the algebra automorphism $\gamma
:e_{k}\rightarrow e_{N-k}$. A similar relation holds for the Hecke algebra
generators via (\ref{hom}). The representation $\mathcal{B}$ of the braid $%
\beta $ in (\ref{beta}) invokes the same automorphism,%
\begin{equation}
b_{j}\beta =\beta b_{N-j},\qquad 1\leq j<N\ ,  \label{betab}
\end{equation}%
which can be verified directly on the abstract algebra level using the
relation%
\begin{equation}
b_{j}\beta _{j-1}\beta _{j}=\beta _{j-1}\beta _{j}b_{1}\ .
\end{equation}%
The last identity is most easily checked graphically by identifying the $%
b_{i}$'s with the generator of Artin's braid group acting on $N$ strings. $%
\beta $ is also invertible, and we can conclude that $C\mathcal{B}^{-1}$
commutes with the Temperley-Lieb action. Hence, we must have%
\begin{equation*}
C\mathcal{B}^{-1}|_{\Gamma _{j}^{(r)}}=\chi _{j}
\end{equation*}%
for some scalar $\chi _{j}\in \mathbb{C}$. This completes the proof.~$%
\blacksquare $\smallskip

In order to completely fix the algebraic expression for $C$ we need to
compute the missing scalar factors $\chi _{j}$ on each invariant subspace $%
\Gamma _{j}^{(r)}$. Due to the fact that $C^{2}=1$ the latter are simply
determined by computing the value of the central element $\mathcal{B}^{2}$
on each $\Gamma _{j}^{(r)}$. In Appendix B, we present this computation for $%
q$ generic on the unrestricted path space $\Gamma _{j}$ using diagrammatic
techniques. Here we argue that this result extends to the root of unity case
as well. \smallskip

\noindent \underline{\textsc{Lemma 5.1}} \textsc{.} \emph{Let }$\beta \in
H_{N}(q)$\emph{\ be defined as in (\ref{beta}). Denote by }$\varrho
_{j}^{(r)}$\emph{\ the irreducible representation given by restricting (\ref%
{pathTL}) to }$\Gamma _{j}^{(r)}$\emph{, i.e. all restricted paths on the
Bratelli diagram ending at }$j$\emph{. Then}%
\begin{equation}
\varrho _{j}^{(r)}(\beta ^{2})=q^{-\frac{N(N-4)}{2}-2j(j+1)}  \label{beta2}
\end{equation}%
\emph{and hence}%
\begin{equation}
\chi _{j}=q^{\frac{N(N-4)}{4}+j(j+1)}  \label{chi}
\end{equation}%
\emph{in Theorem 5.2.}\smallskip

\noindent \textsc{Proof.} We start from the result that $\pi _{j}(\beta
^{2})=q^{-\frac{N(N-4)}{2}-2j(j+1)}$ for generic $q$ on the irreducible
subspace $\Gamma _{j}$; see Appendix B for the proof. Since $\mathcal{B}^{2}$
remains central when taking the limit $q\rightarrow q^{\prime }$ with $%
q^{\prime }=\exp (i\pi /r)$, $r$ integer$~\geq 3$ it suffices to evaluate $%
\mathcal{B}^{2}$ on any path which will belong to the restricted subspace $%
\Gamma _{j}^{(r)}$. Obviously, there always exists such a path, for instance
take the path $\boldsymbol{j}^{\prime }$ from (\ref{jprime}) in the proof of
Theorem 5.1. This fixes $\chi _{j}$ in (\ref{Cbeta}) up to a sign. We take
the same (principal) branch as used in the matrix elements (\ref{etamatrix})
of $\eta $. ~$\blacksquare $\smallskip

For illustration we state in Table 3 below the powers occurring in (\ref%
{beta2}) for some examples.\medskip

\begin{center}
\begin{tabular}{||c||c|c|c|c|c|c|c|c|c|c|c|}
\hline
$N~\backslash ~j$ & 0 & $\frac{1}{2}$ & 1 & $\frac{3}{2}$ & 2 & $\frac{5}{2}$
& 3 & $\frac{7}{2}$ & 4 & $\frac{9}{2}$ & 5 \\ \hline\hline
3 &  & 0 &  & 6 &  &  &  &  &  &  &  \\ \hline
4 & 0 &  & 4 &  & 12 &  &  &  &  &  &  \\ \hline
5 &  & 4 &  & 10 &  & 20 &  &  &  &  &  \\ \hline
6 & 6 &  & 10 &  & 18 &  & 30 &  &  &  &  \\ \hline
7 &  & 12 &  & 18 &  & 28 &  & 42 &  &  &  \\ \hline
8 & 16 &  & 20 &  & 28 &  & 40 &  & 56 &  &  \\ \hline
9 &  & 24 &  & 30 &  & 40 &  & 54 &  & 72 &  \\ \hline
10 & 30 &  & 34 &  & 42 &  & 54 &  & 70 &  & 90 \\ \hline
\end{tabular}%
\medskip

Table 3. Negative powers $y$ of $q$ occurring in the restriction $\beta
^{2}|_{\Gamma _{j}}=q^{-y}$.
\end{center}

\subsection{The case when $q$ is not a root of unity}

Somewhat paradoxically the case when $q$ is on the unit circle but not a
root of unity is simpler from an algebraic point of view and yet the
discussion of quasi-Hermiticity in terms of the path basis becomes more
involved. This is due to the fact that the restriction (\ref{RSOS}) on the
paths on the Bratelli diagram is lifted, and the $q$-integers appearing in
the representation (\ref{pathTL}) can now change sign as the path
progresses. Namely, parametrizing as before $q=\exp (i\pi /r)$ but now with $%
r\in \mathbb{R}$ we have%
\begin{eqnarray}
\lbrack 2j+1]_{q} &>&0,\quad \quad 2\ell r<2j+1<(2\ell +1)r,  \notag \\
\lbrack 2j+1]_{q} &<&0,\quad \quad (2\ell +1)r<2j+1<(2\ell
+2)r,[2j+1]_{q}>0,\quad \ell =0,1,2,...
\end{eqnarray}%
Thus, positivity of the $q$-integers along a path $\boldsymbol{j}\in \Gamma $
is only guaranteed as long as $r>N$. For this segment of the unit circle the
previous constructions and results apply verbatim, with the exception that
the Hamiltonian is diagonalizable without any restriction being placed on
the state space.

\section{Conclusions}

In this paper, we have carried out a detailed and exact analysis of $PT$
symmetry and quasi-Hermiticity for the quantum group symmetric XXZ spin
chain. This model has two key advantages as a laboratory for the in depth
investigation of these ideas: it is finite-dimensional, and it is exactly
solvable. As a consequence of the latter property, there is a well developed
and rich algebraic description of this model. We have used this algebraic
machinery in order to construct an exact expression for the $\eta $
operator, whose key property $\eta H=H^{\ast }\eta $ demonstrates the
quasi-Hermiticity of the model for $q$ a root of unity. In order to develop
this construction, we have been inevitably lead to carry out the procedure
of quantum group reduction. We have thus constructed $\eta $, given by
equation (\ref{eta}), in terms of the path basis in which this reduction is
well defined. Moreover, this construction linked the question of determining
whether $H$ is quasi-Hermitian to the mathematical problem of finding a
self-adjoint representation of the Temperley-Lieb algebra.

Bender and others have introduced the idea of a $C$ operator in the
discussion of PT symmetry \cite{Bender07} which is closely connected with
the notion of quasi-Hermiticity \cite{Most04}. We too have defined such a $C$
operator as $C=P\eta $ ($P$ is the parity operator as discussed in the main
text). This operator is very natural from an algebraic point of view, and
has the realization in terms of the braid $\beta $ given by (\ref{Cbeta}).

Two algebras appear in the description of the quantum spin chain: the
quantum group and the Temperley-Lieb. The quantum version of Schur-Weyl
duality tells us that each is the commutant of the other. The algebraic
construction of $C$ combined with this duality led us naturally to define a
new operator $C^{\prime }=R\eta $ (here $R$ is the spin-reversal operator)
with similar but dual properties to $C$ and with $[C^{\prime },C]=0$. These
properties are summarized in Table 4.{\small \medskip }

We have given a construction of $\eta $, and thus a proof of the reality of
the spectrum of (\ref{H}), that is valid for $q=\exp (i\pi /r)$ for two
regions: $r$ an integer $\geq 3$; and $r>N$ (see Section 5.2). It is
commonly assumed that the spectrum of the model is also real for $q$ on the
unit circle outside of this region, see e.g. \cite{Alc87}. To the best of
our knowledge this assumption is based on numerical investigations of the
Bethe ansatz equations and a rigorous proof of this assertion is missing.
(Obviously, the spectrum is also real when $q\in \mathbb{R}$, for which the
Hamiltonian is Hermitian with respect to the original canonical inner
product on $V^{\otimes N}$.) Clear questions remain as to whether the model
is also quasi-Hermitian in this region, whether there is a clear criterion
that shows this, and if so, whether there is a simple alternative
construction of $\eta $. Preliminary numerical tests which we have carried
out seem to indicate that one in general has to drop the more stringent
condition that $\eta $ intertwines the Temperley-Lieb generators, i.e. $\eta
E_{k}=E_{k}^{\ast }\eta $.

\begin{center}
\begin{tabular}{|c|c|c|c|}
\hline\hline
Operator & Hamiltonian & Temperley-Lieb & Quantum Group \\ \hline\hline
$\eta $ & $\eta H=H^{\ast }\eta $ & $\eta E_{k}=E_{k}^{\ast }\eta $ & $\eta
S^{\pm }=S_{\text{op}}^{\pm }\eta $ \\ \hline
$C=P\eta $ & $[C,H]=0$ & $CE_{k}=E_{N-k}C$ & $[C,S^{\pm }]=[C,S^{z}]=0$ \\ 
\hline
$C^{\prime }=R\eta $ & $[C^{\prime },H]=0$ & $[C^{\prime },E_{k}]=0$ & $%
C^{\prime }S^{\pm }=S^{\mp }C^{\prime },\;C^{\prime }S^{z}=-S^{z}C^{\prime }$
\\ \hline
\end{tabular}%
{\small \medskip }

Table 4. Commutation relations for the operator $\eta $ and the two $C$%
-operators.{\small \medskip }
\end{center}

Finally, we point out that we have also omitted the case $r=2$ or $q=\sqrt{-1%
}$ from our discussion as this case is rather special. The quantum group
reduction as discussed here does not apply to this case and the
corresponding question of quasi-Hermiticity will be investigated in a
separate publication \cite{CK07}.{\small \medskip }

\noindent \textbf{Acknowledgments}. The authors would like to thank Andreas
Fring, Paul Martin, Vladimir Rittenberg and Catharina Stroppel for useful
advice and comments. C.K. is financially supported by a University Research
Fellowship of the Royal Society.

\appendix

\section{Kauffman diagrams \& Levy's reduced words}

In this appendix we review briefly the known graphical calculus associated
with the Temperley-Lieb algebra and give a characterisation of its
irreducible representations which is slightly different from the one in the
main text. We then will employ this graphical calculus in Appendix B to
compute the values of the central element $\beta ^{2}$ defined in equation (%
\ref{beta}) of Theorem 5.2.

We follow ideas put forward by Kauffman \cite{Kauf87}, and we adopt the
conventions of \cite{Levy90} to describe the irreducible representations of $%
TL_{N}(q)$ using primitive left ideals of the Temperley-Lieb algebra. In
what follows we let $q$ be generic, i.e. we treat $q$ as a formal
indeterminate.\medskip

\noindent \underline{\textsc{Definition A.1}} \textsc{[Levy]\cite{Levy90}.} 
\emph{A word }$w=e_{i_{1}}e_{i_{2}}\cdots e_{i_{n}}\in TL_{N}(q)$ \emph{in
the Temperley-Lieb generators is said to possess a \textquotedblleft
jump\textquotedblright\ if the indices of two neighbouring generators differ
by more than one, i.e. }$|i_{k}-i_{k+1}|>1$ \emph{for some} $k=1,...,n-1$%
\emph{.}\medskip

Obviously, the maximum value $k_{\max }$ of jumps which can occur in a word $%
w$ is $k_{\max }=\left\lfloor N/2\right\rfloor -1\,,$ where $\left\lfloor
N/2\right\rfloor $ is the integer part of $N/2$.\medskip

\noindent \underline{\textsc{Definition A.2}} \textsc{[Levy].} \emph{Let }$%
I_{k}\subset TL_{N}(q)$\emph{\ be the left primitive ideal consisting of all
words having at least }$k$\emph{\ jumps. Setting }$I_{-1}\equiv TL_{N}(q)\ $%
\emph{we define for each }$-1<k<k_{\max }$\emph{\ the (vector space) quotient%
}%
\begin{equation}
W_{k}=I_{k}/I_{k+1},  \label{Wk}
\end{equation}%
\emph{i.e. the set of all words which have precisely} $k$ \emph{jumps. For }$%
k=-1$ \emph{we have} $W_{-1}=\mathbb{C}$ \emph{and for }$k=k_{\max }$\emph{\
we set }$W_{k_{\max }}=I_{k_{\max }}$\emph{.}\medskip

Regarded as a vector space, $W_{k}$ is equipped with a natural action of $%
TL_{N}(q)$ and gives rise to an irreducible representation,%
\begin{equation}
\rho _{k}:TL_{N}(q)\rightarrow \limfunc{End}W_{k},\quad a\mapsto \rho
_{k}(a)\qquad \text{with\quad }\rho _{k}(a)w:=aw\;.  \label{rhok}
\end{equation}%
In order to identify the representations $\rho _{k}$ with the correct path
representation $\varrho _{j}$ over $\Gamma _{j}$ given in the main text -
see (\ref{pathTL}) and (\ref{Gammaj}) - we need to compare dimensions. To
this end we follow Levy and introduce for each $W_{k}$ a basis in terms of
the reduced words%
\begin{equation}
w_{n}^{(m)}=e_{m}e_{m-1}\cdots e_{n},\qquad m>n  \label{redw}
\end{equation}%
defining the basis elements to be%
\begin{equation}
w_{m_{1},...,m_{k+1}}=w_{1}^{(m_{1})}w_{3}^{(m_{2})}\cdots
w_{2k+1}^{(m_{k+1})},\qquad 1\leq m_{1}<~...~<m_{k+1}\leq N-1,\qquad
m_{i}>2i-1\ .  \label{wbasis}
\end{equation}%
From this basis definition, one computes the corresponding dimensions to be 
\begin{equation}
\dim W_{k}=\binom{N-1}{k+1}-\binom{N-1}{k-1}\ .
\end{equation}%
Upon comparing this result with the multiplicity formula (\ref{dimGamj}) on
the Bratelli-diagram one can conclude that the following representations are
isomorphic%
\begin{equation}
W_{k}\cong \Gamma _{j=N/2-k-1}\ .  \label{iso}
\end{equation}%
Our motivation for introducing the representations $\rho _{k}$ is that they
allow for a convenient graphical calculus. For given $k$, each basis element
in (\ref{wbasis}) corresponds to a diagram of $k+1$ (possibly nested) caps
and $N-2(k+1)$ vertical lines on which the Temperley-Lieb generators $e_{i}$
act in a simple manner. We demonstrate this for a simple example to keep
this paper self-contained.

\subsection{Example $N=6$}

Let the identity element in $TL_{N}(q)$ correspond to a diagram consisting
of $N=6$ strands and each Temperley-Lieb generator $e_{k}$ be represented by
a diagram similar to the one for $e_{3}$ depicted in the figure below,%
\begin{equation}
\includegraphics[scale=0.8]{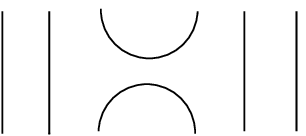}
\end{equation}
The algebra multiplication corresponds to composing diagrams from below. The
basis elements (\ref{wbasis}) spanning $W_{k}$ should be considered as
equivalence classes of words. By abuse of notation we denote them by the
same symbols as the algebra elements. Since $I_{k}$ is a left ideal and the
algebra action in (\ref{rhok}) is defined to be from the left it suffices to
depict the elements (\ref{wbasis}) by the bottom half of their respective
diagrams. For instance for $N=6$ and $k=k_{\max }=2$ we have the five words%
\begin{equation}
e_{1}e_{3}e_{5}= \includegraphics[scale=0.5]{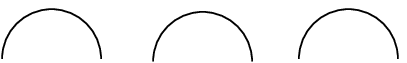} \quad ,\qquad
e_{1}~e_{4}e_{3}~e_{5}= \includegraphics[scale=0.5]{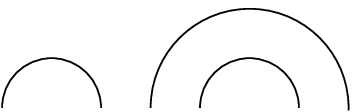}\ ,
\end{equation}%
\begin{equation}
e_{2}e_{1}~e_{3}~e_{5}= \includegraphics[scale=0.5]{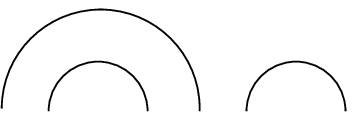}\ ,\qquad
e_{2}e_{1}~e_{4}e_{3}~e_{5}= \includegraphics[scale=0.5]{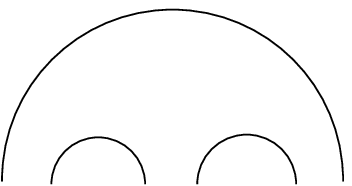}\ ,
\end{equation}%
and%
\begin{equation}
e_{3}e_{2}e_{1}~e_{4}e_{3}~e_{5}= \includegraphics[scale=0.5]{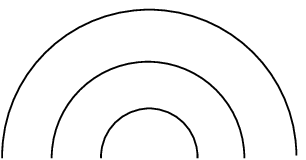}\
.
\end{equation}%
The action of the Temperley-Lieb algebra on these 5 diagrams from below
leads to simple permutations of the basis elements. For example acting with $%
e_{2}$ on the first diagram we obtain,%
\begin{equation}
\includegraphics[scale=0.5]{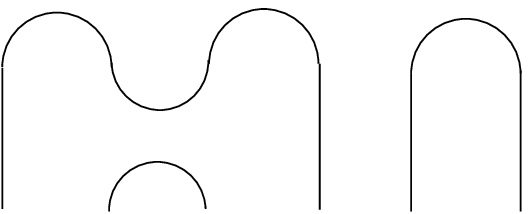}\quad =\quad %
\includegraphics[scale=0.5]{n6basis3.eps}\ .
\end{equation}%
Any closed loops occurring in this process yield factors $-(q+q^{-1})$
according to the defining relation $e_{i}^{2}=-(q+q^{-1})e_{i}$.

\section{Computation of $\protect\beta ^{2}$}

We now turn to the graphical computation of the central element $\beta
^{2}\in H_{N}(q)$ involving the braid $\beta $ defined in (\ref{beta}). The
Hecke algebra generators also have a graphical representation. Namely, each
generator $b_{i}$ acts on the identity diagram of $N$ parallel strands by
crossing the $i^{\text{th}}$ strand over the ($i+1$)$^{\text{th}}$ one. For $%
b_{i}^{-1}$ the ($i+1$)$^{\text{th}}$ strand is on top. The braid $\beta $
has then the following graphical depiction%
\begin{equation}
\includegraphics[scale=0.8]{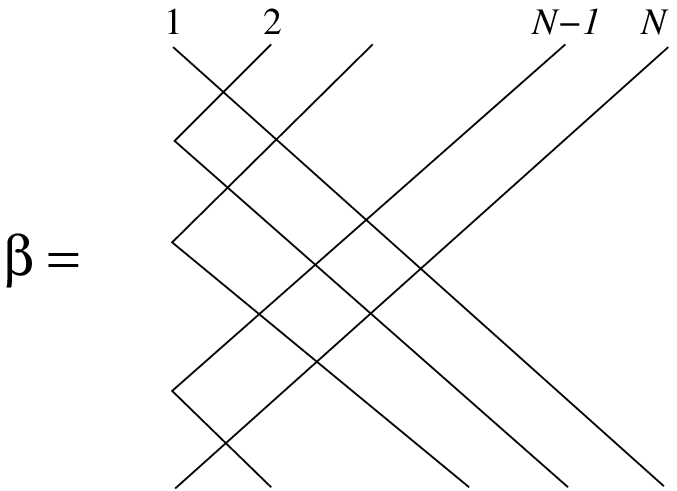}
\end{equation}%
where the NW-SE lines cross over the NE-SW lines. In order to compute the
action of $\beta ^{2}$ on the basis elements in $W_{k}\cong \Gamma
_{j=N/2-k-1}$ we need further graphical rules. All of them are a direct
consequence of the homomorphism (\ref{hom}) which graphically amounts to the
following equality of diagrams,%
\begin{equation}
\includegraphics[scale=0.7]{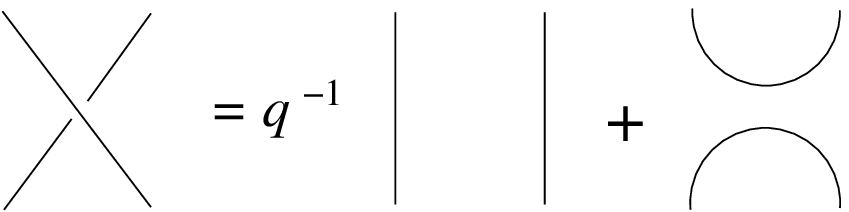}  \label{rule0}
\end{equation}%
An analogous picture holds for the relation $b_{i}^{-1}\rightarrow q+e_{i}$.
Repeated use of these relations together with the defining relations of the
Temperley-Lieb generators now yields the following\medskip

\noindent \underline{\textsc{Lemma B.1}}\textsc{.} \emph{Employing the
identification (\ref{hom}) one verifies the following identities}

\begin{itemize}
\item[(i)] $b_{i}e_{i}=-qe_{i}$\emph{; compare with the depiction below.}%
\begin{equation}
\includegraphics[scale=0.8]{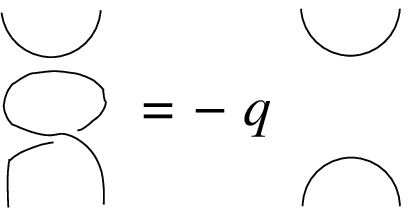}
\end{equation}

\item[(ii)] $b_{i}b_{i\pm 1}e_{i}=q^{-1}e_{i\pm 1}e_{i}$\emph{; see the
pictures below.}%
\begin{equation}
\includegraphics[scale=0.7]{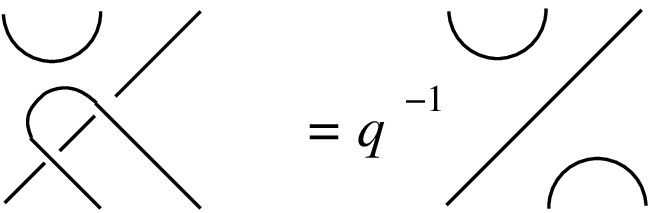}\qquad %
\includegraphics[scale=0.7]{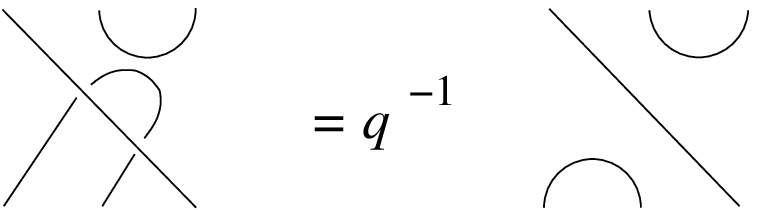}
\end{equation}
\end{itemize}

\noindent \textsc{Proof.} A trivial computation, that follows either
directly from the identification (\ref{hom}), or from the graphical rule (%
\ref{rule0}) together with the identification of a complete circle with the
coefficient $-(q+q^{-1}).$ $\blacksquare $\medskip

We are now in a position to derive the values of the central element $\beta
^{2}$ in the irreducible representation (\ref{pathTL}) over the path space $%
\Gamma _{j}$.\medskip

\noindent \underline{\textsc{Lemma B.2}}\textsc{.} \emph{Assume }$q$ \emph{%
to be generic and} \emph{let }$\beta \in H_{N}(q)$\emph{\ be defined as in (%
\ref{beta}), i.e.}%
\begin{equation*}
\beta =\beta _{1}\beta _{2}\cdots \beta _{N-1},\qquad \beta
_{n}=b_{n}b_{n-1}\cdots b_{1}\ .
\end{equation*}%
\emph{\ Denote by }$\varrho _{j}$\emph{\ the irreducible representation
given by (\ref{hom}) and restricting (\ref{pathTL}) to }$\Gamma _{j}$\emph{,
i.e. all paths on the Bratelli diagram ending at }$j$\emph{. Then}%
\begin{equation}
\varrho _{j}(\beta ^{2})=q^{-\frac{N(N-4)}{2}-2j(j+1)}\ .
\end{equation}%
\medskip

\noindent \textsc{Proof.} According to our remarks in Appendix A we can
exploit the fact that for generic $q$ the finite-dimensional irreducible
representations of $TL_{N}(q)$ respectively $H_{N}(q)$ are determined by
their dimensions up to isomorphism. Since $\beta ^{2}$ is central its value
does not depend on the particular choice of the representation as long as we
stay in the same isomorphism class. We can therefore identify $W_{k}\cong
\Gamma _{j=N/2-k-1}$ and compute the action of $\beta $ on the reduced words
in $W_{k},$ where it is particularly simple. According to Schur's lemma it
does not matter which word we use and we focus our attention on the word $%
w_{1,3,5,...,2k+1}$ depicted below,%
\begin{equation}
\includegraphics[scale=0.7]{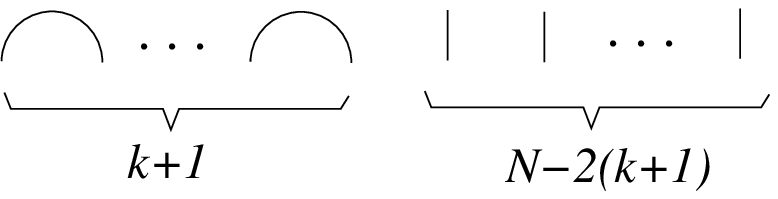}
\end{equation}%
Acting with $\beta $ on this word from below we obtain the diagram shown here%
\begin{equation}
\includegraphics[scale=0.8]{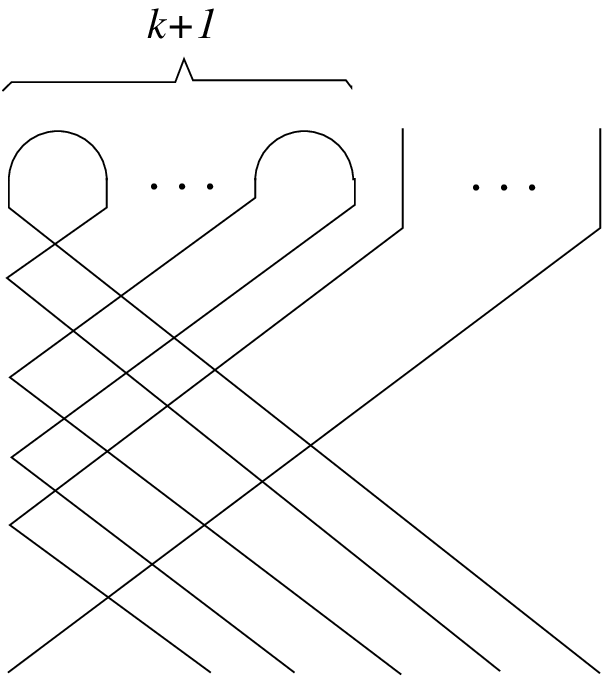}
\end{equation}%
We will now undo this braid in a number of successive steps using the
diagrammatic rules of the preceding lemma. We start with the left most cap.
Employing rule (i) of the preceding lemma we untwist it once and obtain a
factor $-q$. Applying rule (ii) from the previous lemma we pull it over $N-2$
NE-SW lines producing the factor $-q^{3-N}$. The resulting diagram is
depicted below.%
\begin{equation}
\includegraphics[scale=0.8]{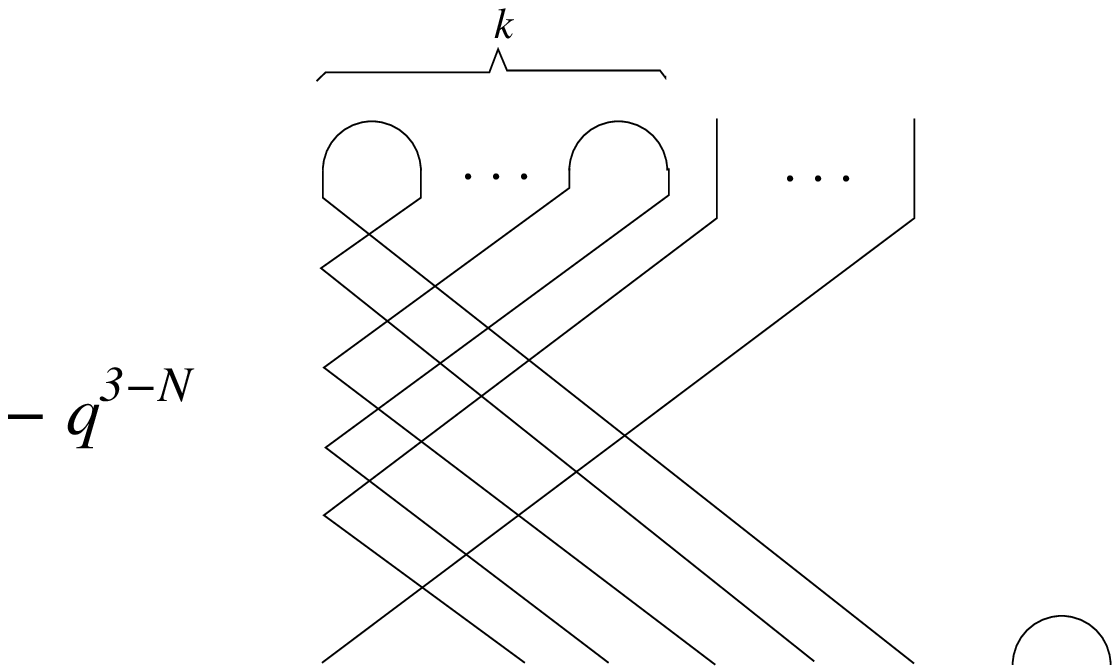}
\end{equation}%
Thus, we are back at the starting point, but now with a diagram which has
one less cap. Repeating the same steps as before we end up with the diagram 
\begin{equation}
\includegraphics[scale=0.8]{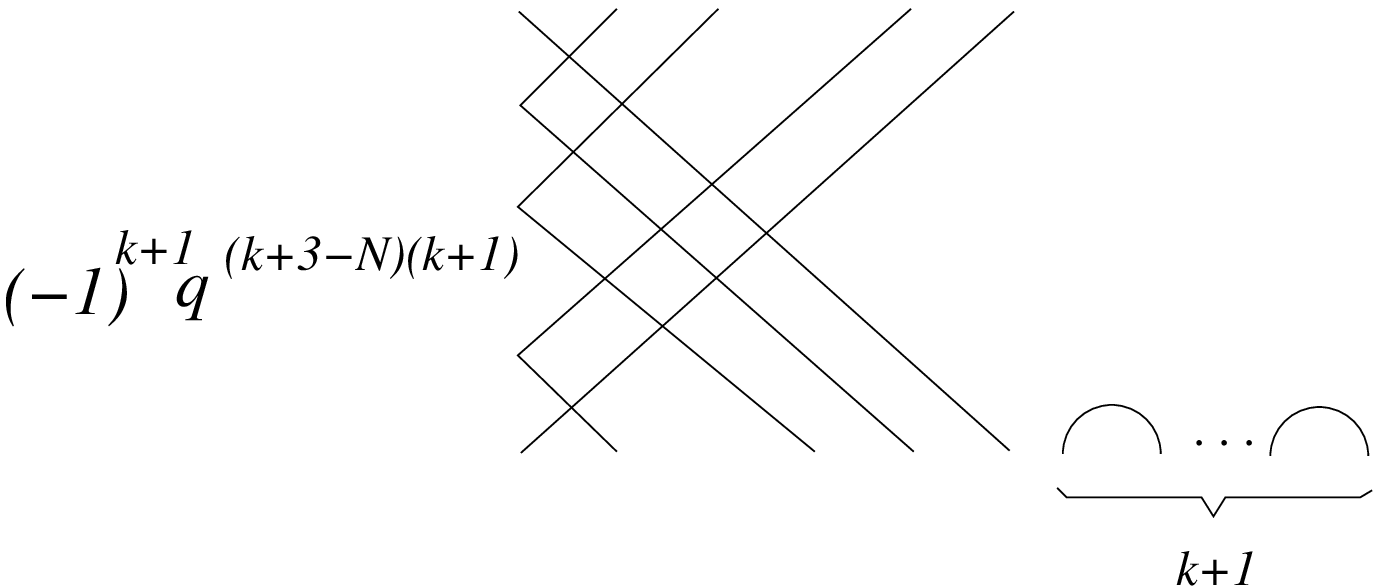}  \label{nocaps}
\end{equation}%
from which all caps are removed. In the remaining graph we need to undo all
the crossings of lines according to (\ref{rule0}). However, we can discard
all the terms which introduce additional caps since according to (\ref{Wk})
these are identified with zero under the quotient. Thus, each of the $%
(N-2k-3)(N-2k-2)/2$ vertices in the diagram (\ref{nocaps}) yields a factor $%
q^{-1}$ and we end up with the diagram%
\begin{equation}
(-1)^{k+1}q^{-N(N-2k+3)/2-k(k+1)}\quad %
\includegraphics[scale=0.8]{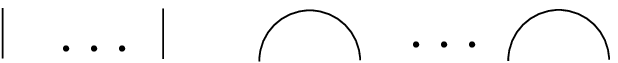}  \label{betaonce}
\end{equation}%
Up to a factor, we have simply obtained the diagram reflected about the
vertical axis. By reflection symmetry we deduce that applying $\beta $ twice
simply produces the $q$-factor in (\ref{betaonce}) to the power two. Upon
replacing $k+1=N/2-j$ according to (\ref{iso}) we obtain the desired result. 
$\blacksquare $

\section{Derivation of the Clebsch-Gordan coefficients}

In this section, we derive the Clebsch-Gordan coefficients (\ref{cgc1}), (%
\ref{cgc2}) entering the definition of the path basis vectors (\ref{path
state}) according to (\ref{path matrix}). We use the convention (\ref{spin j
rep}) for the irreducible representations of the quantum group $%
U_{q}(sl_{2}) $. We include these details as some derivations in the
literature are known to contain minor errors.

\subsection{The CG coefficients for $J=j+1/2,~j_{1}=j,~j_{2}=1/2$}

We start with the representation $\pi _{j+1/2}\subset \pi _{j}\otimes \pi
_{1/2}$ and take as highest weight vector%
\begin{equation}
\left\vert J,M=J\right\rangle =\left\vert j,j\right\rangle \otimes
\left\vert \tfrac{1}{2},\tfrac{1}{2}\right\rangle \ .
\end{equation}%
Successive action with $\Delta (S^{-})$ on the highest weight vector yields
the expression%
\begin{eqnarray}
\Delta (S^{-})^{m}\left\vert J,J\right\rangle &=&q^{j-\frac{m-1}{2}%
}[m]\left( \frac{[2j]![m-1]!}{[2j-m+1]!}\right) ^{\frac{1}{2}}\left\vert
j,j-m+1\right\rangle \otimes \left\vert \tfrac{1}{2},-\tfrac{1}{2}%
\right\rangle  \notag \\
&&+q^{-\frac{m}{2}}\left( \frac{[2j]![m]!}{[2j-m]!}\right) ^{\frac{1}{2}%
}\left\vert j,j-m\right\rangle \otimes \left\vert \tfrac{1}{2},\tfrac{1}{2}%
\right\rangle  \notag \\
&=&\left( \frac{[2J]![m]!}{[2J-m]!}\right) ^{\frac{1}{2}}\left\vert
J,J-m\right\rangle =\left( \frac{[2j+1]![m]!}{[2j-m+1]!}\right) ^{\frac{1}{2}%
}\left\vert J,J-m\right\rangle \ .
\end{eqnarray}%
From this formula we infer the first identity (\ref{cgc1}) for the CG
coefficients.

\subsection{The CG coefficients for $J=j-1/2,\ j_{1}=j,\ j_{2}=1/2$}

We now turn to the representation $\pi _{j-1/2}\subset \pi _{j}\otimes \pi
_{1/2}$. The highest weight vector is now determined by the relation (the
factor $q^{-1/2}$ is introduced to obtain a more symmetric expression for
the CG coefficients)%
\begin{equation}
0=\Delta (S^{+})\left\vert J,J\right\rangle =\Delta (S^{+})\left\{ q^{-\frac{%
1}{2}}\left\vert j,j\right\rangle \otimes \left\vert \tfrac{1}{2},-\tfrac{1}{%
2}\right\rangle +\gamma q^{-\frac{1}{2}}\left\vert j,j-1\right\rangle
\otimes \left\vert \tfrac{1}{2},\tfrac{1}{2}\right\rangle \right\}
\end{equation}%
and one easily finds%
\begin{equation}
\gamma =-\frac{q^{j+\frac{1}{2}}}{[2j]^{\frac{1}{2}}}\ .
\end{equation}%
In a similar manner as above, one proves by induction that%
\begin{eqnarray}
\Delta (S^{-})^{m}\left\vert J,J\right\rangle &=&q^{-\frac{m+1}{2}%
}[2j-m]\left( \frac{[2j-1]![m]!}{[2j][2j-m]!}\right) ^{\frac{1}{2}%
}\left\vert j,j-m\right\rangle \otimes \left\vert \tfrac{1}{2},-\tfrac{1}{2}%
\right\rangle  \notag \\
&&-q^{j-\frac{m}{2}}\left( \frac{[2j-1]![m+1]!}{[2j][2j-m-1]!}\right) ^{%
\frac{1}{2}}\left\vert j,j-m-1\right\rangle \otimes \left\vert \tfrac{1}{2},%
\tfrac{1}{2}\right\rangle  \notag \\
&=&\left( \frac{[2J]![m]!}{[2J-m]!}\right) ^{\frac{1}{2}}\left\vert
J,J-m\right\rangle =\left( \frac{[2j-1]![m]!}{[2j-m-1]!}\right) ^{\frac{1}{2}%
}\left\vert J,J-m\right\rangle \ .
\end{eqnarray}%
From this result we read off 
\begin{equation}
\left\vert 
\begin{array}{ccc}
j & \frac{1}{2} & j-\frac{1}{2} \\ 
m & \alpha & m+\alpha%
\end{array}%
\right\vert _{q}=-2\alpha q^{\alpha (j+1)+\frac{m}{2}}\left( \frac{%
[j-2\alpha m]}{[2j]}\right) ^{\frac{1}{2}}
\end{equation}%
In order to have the crucial identity%
\begin{equation}
\sum_{m,\alpha }\left\vert 
\begin{array}{ccc}
j & \frac{1}{2} & j^{\prime } \\ 
m & \alpha & m^{\prime }%
\end{array}%
\right\vert _{q}\left\vert 
\begin{array}{ccc}
j & \frac{1}{2} & j^{\prime \prime } \\ 
m & \alpha & m^{\prime \prime }%
\end{array}%
\right\vert _{q}=\delta _{j^{\prime },j^{\prime \prime }}\delta _{m^{\prime
},m^{\prime \prime }}
\end{equation}%
we renormalize the CG coefficients by the factor $-\sqrt{[2j+1]/[2j]}$ such
that we obtain the second identity (\ref{cgc2}).

\subsection{Identities involving the CG coefficients}

For the derivation of (\ref{pathTL}) one requires the following set of
identities, which we list without proof.%
\begin{equation*}
\left\vert 
\begin{array}{ccc}
j\pm \frac{1}{2} & \frac{1}{2} & j \\ 
m+\alpha & -\alpha & m%
\end{array}%
\right\vert _{q}=\mp 2\alpha q^{-\alpha }\sqrt{\frac{[2j+1]}{[2j+1\pm 1]}}%
\left\vert 
\begin{array}{ccc}
j & \frac{1}{2} & j\pm \frac{1}{2} \\ 
m & \alpha & m+\alpha%
\end{array}%
\right\vert _{q}
\end{equation*}%
\begin{multline*}
\left\vert 
\begin{array}{ccc}
j & \frac{1}{2} & j\pm \frac{1}{2} \\ 
m & \alpha & m+\alpha%
\end{array}%
\right\vert _{q}\left\vert 
\begin{array}{ccc}
j\pm \frac{1}{2} & \frac{1}{2} & j\pm 1 \\ 
m+\alpha & -\alpha & m%
\end{array}%
\right\vert _{q}= \\
q^{2\alpha }\left\vert 
\begin{array}{ccc}
j & \frac{1}{2} & j\pm \frac{1}{2} \\ 
m & -\alpha & m-\alpha%
\end{array}%
\right\vert _{q}\left\vert 
\begin{array}{ccc}
j\pm \frac{1}{2} & \frac{1}{2} & j\pm 1 \\ 
m-\alpha & \alpha & m%
\end{array}%
\right\vert _{q}
\end{multline*}%
\begin{multline*}
\left\vert 
\begin{array}{ccc}
j & \frac{1}{2} & j\pm \frac{1}{2} \\ 
m & \alpha & m+\alpha%
\end{array}%
\right\vert _{q}\left\vert 
\begin{array}{ccc}
j\pm \frac{1}{2} & \frac{1}{2} & j \\ 
m+\alpha & -\alpha & m%
\end{array}%
\right\vert _{q}= \\
-q^{-2\alpha (2j+1)}\frac{[j\pm 2\alpha m+\frac{1\pm 1}{2}]}{[j\mp 2\alpha m+%
\frac{1\pm 1}{2}]}\left\vert 
\begin{array}{ccc}
j & \frac{1}{2} & j\pm \frac{1}{2} \\ 
m & -\alpha & m-\alpha%
\end{array}%
\right\vert _{q}\left\vert 
\begin{array}{ccc}
j\pm \frac{1}{2} & \frac{1}{2} & j \\ 
m-\alpha & \alpha & m%
\end{array}%
\right\vert _{q}
\end{multline*}%
\begin{equation*}
\sum_{j^{\prime }=j\pm 1/2}[2j^{\prime }+1]^{\frac{1}{2}}\left\vert 
\begin{array}{ccc}
j & \frac{1}{2} & j^{\prime } \\ 
m & \alpha & m+\alpha%
\end{array}%
\right\vert _{q}\left\vert 
\begin{array}{ccc}
j^{\prime } & \frac{1}{2} & j \\ 
m+\alpha & \alpha & m+2\alpha%
\end{array}%
\right\vert _{q}=0
\end{equation*}%
\begin{equation*}
\sum_{j^{\prime }=j\pm 1/2}\pm \lbrack 2j^{\prime }+1]^{\frac{1}{2}%
}\left\vert 
\begin{array}{ccc}
j & \frac{1}{2} & j^{\prime } \\ 
m & \alpha & m+\alpha%
\end{array}%
\right\vert _{q}\left\vert 
\begin{array}{ccc}
j^{\prime } & \frac{1}{2} & j \\ 
m+\alpha & -\alpha & m%
\end{array}%
\right\vert _{q}=-2\alpha q^{-\alpha }[2j+1]^{\frac{1}{2}}
\end{equation*}

\end{document}